\documentclass[reprint,amsmath,amssymb,aps, prb]{revtex4-2}
\usepackage{hyperref}
\usepackage[english]{babel} 
\usepackage{amssymb}
\usepackage{amsmath}
\usepackage{txfonts}
\usepackage{mathdots}
\usepackage[classicReIm]{kpfonts}
\usepackage{graphicx}
\usepackage[dvipsnames]{xcolor}
\def\arrvline{\hfil\kern\arraycolsep\vline\kern-\arraycolsep\hfilneg}
\usepackage{dcolumn}
\usepackage{bm}

\begin{document}

\title{Warm-up spectroscopy of quadrupole-split nuclear spins in \textit{n}-GaAs epitaxial layers }

\author{V. M. Litvyak}
\affiliation{St. Petersburg State University, Ulyanovskaya 1, St.~Petersburg 198504, Russia}
\author{R. V. Cherbunin}
\affiliation{St. Petersburg State University, Ulyanovskaya 1, St.~Petersburg 198504, Russia}
\author{V. K. Kalevich}
\affiliation{Ioffe Institute, Politekhnicheskaya 26, St.~Petersburg 194021, Russia}
\author{A. I. Lihachev}
\affiliation{Ioffe  Institute, Politekhnicheskaya 26, St.~Petersburg 194021, Russia}
\author{A. V. Nashchekin}
\affiliation{Ioffe  Institute, Politekhnicheskaya 26, St.~Petersburg 194021, Russia}
\author{M. Vladimirova}
\affiliation{Laboratoire Charles Coulomb, UMR 5221 CNRS/Universit\'{e} de Montpellier, France}
\author{K. V. Kavokin}
\affiliation{St.~Petersburg State University, Ulyanovskaya 1, St. Petersburg 198504, Russia}

\begin{abstract}
The efficiency of the adiabatic demagnetization of nuclear spin system (NSS) of a solid is limited, if quadrupole effects are present. Nevertheless, despite a considerable quadrupole interaction, recent experiments validated the thermodynamic description of the NSS in GaAs.  
This suggests that nuclear spin temperature can be used as a probe of nuclear magnetic resonances. We implement this idea by analyzing the modification of the NSS temperature in response to an oscillating magnetic field at various frequencies, an approach termed as the warm-up spectroscopy. 
It is tested in a $n$-GaAs sample where both mechanical strain and built-in electric field may contribute to the quadrupole splitting, yielding the parameters of electric field gradient tensors for $^{75}$As and both Ga isotopes, $^{69}$Ga and $^{71}$Ga.
\end{abstract}

\date{\today}
\maketitle
\section{Introduction }

The nuclear spin system (NSS) in semiconductors is thermally isolated from its environment, because the thermodynamic equilibrium within the NSS is reached on the $\mathrm{\sim}$100 µs scale, much faster than the equilibrium between the NSS and the lattice. The latter is established by spin-lattice relaxation, which takes seconds and even minutes \cite{OO1,PhysRevB.94.081201,PhysRevB.95.125312}. This means, for instance, that while the spin polarization of nuclei might be short-living in weak or zero magnetic field, the energy accumulated in spin degrees of freedom is stored for much longer times. This hierarchy of relaxation times lies at the base of the powerful theoretical concept of the spin temperature \cite{AbragamProctor}, which is assumed to be the sole characteristic of the NSS state on the timescale longer than 1ms. This concept is particularly fruitful at  vanishing magnetic fields, where various magnetically ordered states have been predicted to emerge if the NSS temperature is made sufficiently low \cite{Merkulov1982,Merkulov1987,Merkulov1998,vladimirova2021}.

\begin{figure}[!h]
	\includegraphics[width=3.4in]{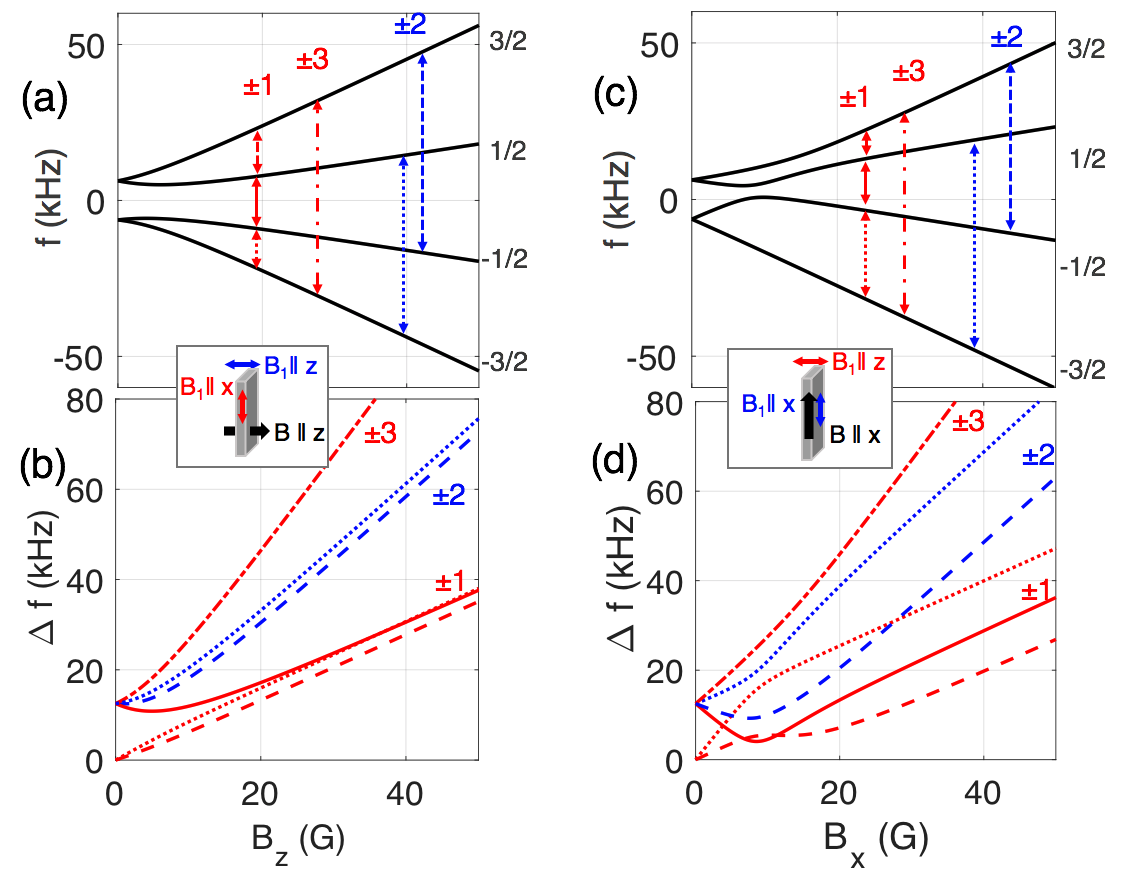}
	\caption{ Spin states (a, c) and OMF-induced spin transition frequencies (b, d) in the presence of Zeeman and quadrupole interaction (calculated from Eq.~\ref{GrindEQ__7_} using parameters in Table 1  for $^{75}$As). Magnetic field is directed along the growth axis (a-b) and in the plane (c-d). Arrows indicate  spin transitions shown below with the same color code. In orthogonal geometries ($B \perp B_1$) spin projection may change by $\pm1$, $\pm3$ (red), in parallel geometries ($B \parallel B_1$)  by $\pm2$ (blue). The notations $\pm 1/2$, $\pm 3/2$ correspond to the spin projections in the high-field limit.}
	\label{fig:fig1}
\end{figure}

One of the most efficient strategies to cool down the NSS in a semiconductor is to combine (i) optical orientation of electrons that polarize NSS via hyperfine interaction under longitudinal magnetic field and (ii) subsequent adiabatic demagnetization to zero field \cite{OO1,PhysRevLett.20.491,PhysRevB.94.081201,PhysRevB.95.125312,JETP1982,CH11Spin2017,PhysRevB.97.041301}.
Adiabatic demagnetisation is a key ingredient of this strategy, because it allows one to reach record low temperatures of the order of few microkelvin in \textit{n}-GaAs \cite{JETP1982,PhysRevB.97.041301,kotur2021deep}, close to fractions of microkelvin required to realize nuclear magnetic ordering \cite{Merkulov1982,Merkulov1987,Merkulov1998,vladimirova2021}.

The lowest spin temperature ${\theta }^{ad}_N$ that can be reached using adiabatic demagnetization to zero field, if the NSS has been preliminary cooled to the temperature ${\theta }_N$ in the magnetic field $B$, is limited by the heat capacity of the nuclear spin reservoir. The heat capacity can be decribed in terms of the local nuclear field $B_L$ that characterises all nuclear spin interactions excluding the Zeeman interaction. During adiabatic demagnetization of the NSS from the magnetic field $B$ to zero field, the spin temperature is reduced by the factor \cite{CH11Spin2017,Goldman}:
\begin{equation} \label{GrindEQ__1_} 
\theta_N / \theta_N^{ad} = \sqrt{(B^2+B^2_L)/B^2_L}.
\end{equation} 
One can see from Eq.~(\ref{GrindEQ__1_}) that the efficiency of cooling is limited by the local field. 
Since all GaAs isotopes have quadrupole moments,  electric field gradients (EFG) induced by the lattice deformations \cite{doi:10.1063/1.1923191} or electric fields \cite{PhysRevB.67.085308} split nuclear spin levels as illustrated in Fig.~\ref{fig:fig1} and increase local fields substantially.
Quadrupole effects in heterostructures like quantum dots \cite{Nat09,Chekhovich2012,Chekhovich2017}, quantum wells \cite{PhysRevLett.72.1368} or microcavities \cite{PhysRevB.97.041301}, and even in epitaxial layers \cite{Lit18} have been reported. In certain quantum dots they are so strong, that NSS fails to thermalize, and the concept of the nuclear spin temperature loses its sense and validity \cite{Nat09}.  By contrast, in "unstrained" GaAs/(Al,Ga)As quantum dots,  quantum wells, microcavities and epitaxial layers, which are characterized by a weaker quadrupole splitting, the thermalization has been demonstrated \cite{PhysRevB.97.041301,Chekhovich2012,Chekhovich2017,PhysRevLett.72.1368,kotur2021deep}. Nevertheless, even in epilayers of \textit{n}-GaAs local fields are of order of $B_L \approx 8$~G, i.e. approximately $5$ times larger than local field due to dipole-dipole interactions $B_{dd}=1.5$~G \cite{PhysRevB.15.5780,PhysRevB.97.041301,Lit18,kotur2021deep}. This fact can be explained by sample deformation during its mounting and further cooling, or by electric field from surface charges or charged impurities \cite{RSI08}.

Motivated by the validity of the thermodynamic description of the NSS in presence of weak quadrupole effects, we propose to use nuclear spin temperature as a sensitive probe of the quadrupole effects. Our method can be considered as an implementation  of the optically detected nuclear magnetic resonance (NMR) spectroscopy, where NMR is probed via changes of the NSS temperature, or in other words via its warm up by oscillating magnetic fields (OMF). The "warm-up" approach to NMR detection is essential in external static fields weaker than the local field (including the case of zero external field), where spin temperature is the unique way to characterize the NSS \cite{AbragamProctor}.

The experimental method that we propose comprises (i) optical cooling  and demagnetization of the  NSS down to a selected magnetic field $B$, (ii) exciting  it by a weak OMF $B_1$ in the frequency range from zero to radiofrequencies (RF), and (iii) measuring the rate at which NSS temperature changes in the presence of the OMF. This warm-up rate is determined immediately after heating by measuring the NSS magnetization in a very low probe magnetic field $\approx 1$~G, for which purpose the Hanle effect under optical orientation of electron spins is used \cite{Lit18,JETP99,Bill83}.
 Repeating the entire procedure with different OMF frequencies, we end up with the warm-up spectrum at a given static field $B$. It provides us with the information on the spin  transition frequencies of different isotopes and their relative probabilities.   

First experiments of this kind were performed back in 1980s in epilayers of \textit{n}-GaAs in zero magnetic field \cite{Bill83}, see also the review \cite{CH11Spin2017}.
However, the resulting spectra could not be understood.
%
In this work,  we apply warm-up spectroscopy under various orientations and intensities of the static magnetic field $B$, as well as different orientations of the OMF. This  allows us to address the ensemble of the quadrupole-split nuclear spin transitions, as illustrated  in Fig.~\ref{fig:fig1}. 
%

The parameters relevant for quadrupole effects are determined from the analysis of the  experimental data  in the framework of the model accounting for both Zeeman interaction of the nuclear magnetic dipole with the static magnetic field $B$ and the interaction of the nuclear electric quadrupole moment $Q$ with EFGs. This allows us to identify the contribution of different isotopes in the spectra and unravel in-plane and longitudinal components of the EFG tensor. Then, three possible sources of the EFGs are considered: (i) an uniaxial deformation either along the growth axis or in the plane, (ii) a shear strain in the plane, (iii) a static electric field perpendicular to the sample surface. 
However, characterization of the sample by electron beam induced current (EBIC) technique shows that the latter contribution is small enough compared to the strain effects and can be neglected. 
Thus EFG tensor is uniquely determined by the strain tensor and a fourth rank gradient-elastic tensor $S_{ijkl}$. The latter can be characterized by only two components, $S_{11}$ and $S_{44}$, because GaAs is a cubic crystal.
Our results suggest that  the ratio of diagonal and off-diagonal components of the gradient-elastic tensor $S_{11}/S_{44}$ for $^{75}$As reported recently in Ref.~\onlinecite{Griffiths2019} must be revisited. 

The paper is organized as follows. The next Section presents the sample and experimental setup. It is followed by the presentation of the experimental protocol of the warm-up spectroscopy. Section IV is devoted to the experimental results, Section V to the model, and Section VI to the interpretation of the results in its light. Section VII summarizes and concludes the paper, {while EBIC characterization results are presented in Appendix, Section \ref{sec:appendix}}.

\section{Samples and experimental setup}
\begin{figure}
	\includegraphics[width=3.1in]{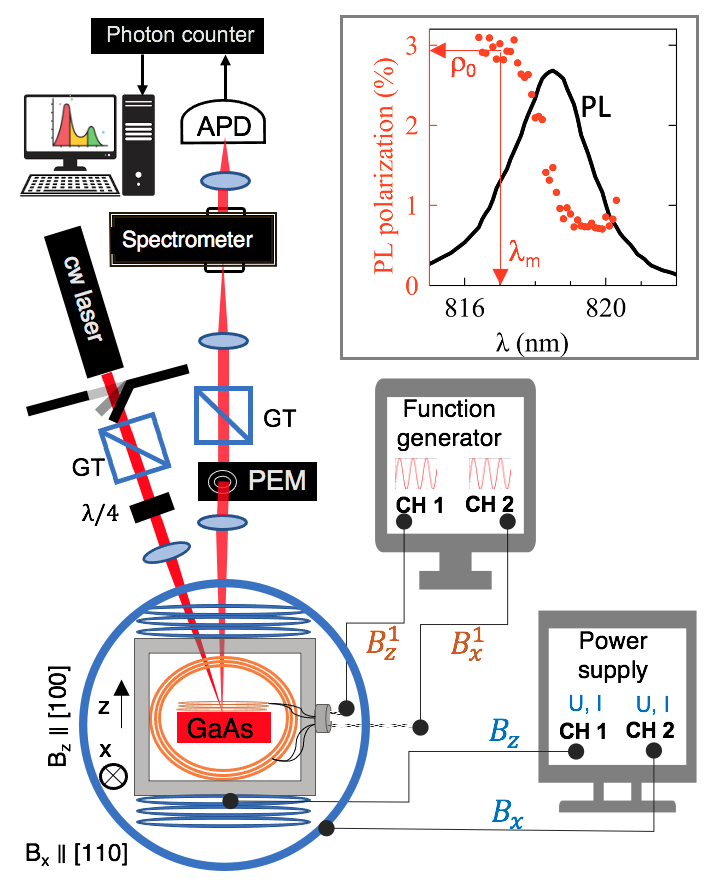}
	\caption{Sketch of the experimental set-up. APD is the avalanche photodiode. GT is the Glan-Taylor prism. PEM is the photo-elastic modulator. Inset shows  spectra of PL  (black line) and PL polarization degree (red points) for \textit{n}-GaAs sample studied in this paper. Red arrows point the PL polarization degree ${\rho}_{0}$ measured at the wavelength $\lambda_m$=817 nm chosen for detection of the NSS warm-up rates.}
	\label{fig:setup}
\end{figure}
We study the same GaAs layer as in Refs.~\onlinecite{Lit18,Bill83}. The details of the growth procedure are described in Ref.~\onlinecite{Sem89}. The layer of $77$~$\muup$m is grown by the liquid phase epitaxy on a \textit{p}-GaAs [100] substrate. The nominal donor concentration in the close-to-the-surface area addressed in PL experiments is $n_d \approx  10^{15}$~cm$^{-3}$. The half-width at half-maximum of the Hanle curve measured at power density $P=200$~W/cm$^{2}$ used in the warm-up experiments is $B_{1/2}=10$~G. This  indicates electron spin relaxation time $T_s \approx 20$~ns, consistent with nominal donor density \cite{PhysRevB.66.245204}. Nuclear spin-lattice relaxation time 
$T_1 \approx 80$~s was measured at lattice temperature $T_L=20$~K and at $B_x=1$~G  by the technique of Ref.~\onlinecite{PhysRevB.94.081201}. 

The experimental setup is sketched in Fig.~2. The sample is cooled down to $T_L=20$~K in a closed-cycle cryostat. The cryostat cold finger is positioned inside two pairs of orthogonal coils, that create magnetic fields oriented along the growth axis ($B_z \parallel [100]$), and in the plane of the sample ($B_x \parallel [011]$). Optical pumping and PL excitation are realized with a laser diode emitting at $\lambda=780$~nm. After passing through a linear polarizer (a Glan-Taylor prism) and a quarter-wave plate, the light beam of $15$~mW power is focused on a $100$~$\muup$m diameter spot on the sample surface. 

A typical PL spectrum and the corresponding PL polarization degree measured with a $0.55$-m spectrometer coupled to an avalanche photodiode are shown in Fig.~\ref{fig:setup}, inset. One can see that the PL spectrum is rather broad, corresponding to overlapping emission of the free excitons, excitons bound to neutral donors (D$^{0}$X), and excitons bound to charged donors (D$^{+}$X). The polarization degree is the highest at the short-wavelength shoulder of the PL line \cite{PhysRevB.66.245204}. Therefore, in our  measurements of the nuclear spin warm-up rate  we choose $\lambda=\lambda_m=817$~nm to monitor  PL polarization induced by the Overhauser field $B_N$. This ensures an optimum trade-off between PL intensity and polarization degree. At $B=0$ and in the absence of the nuclear spin polarization  we get  PL  polarization degree $\rho_0 \approx 3$~\% at $\lambda=\lambda_m$.

The polarized PL obtained within the warm-up spectroscopy protocol (see next Section) is modulated at $50$~kHz by a photoelastic modulator (PEM) and spectrally dispersed by the spectrometer. The PL signal at $\lambda_m$ is detected by a silicon avalanche photodiode (APD) connected to a two-channel photon counter synchronized with the PEM.  Three pairs of Helmholtz coils compensate for the laboratory fields with $0.005$~G precision (not shown in Fig.~\ref{fig:setup}). Two small coils, each comprising ten turns of the enameled copper wire, are placed inside the cryostat near the sample in order to apply OMF in the frequency range from zero to hundreds of kHz either along the growth axis ($B^1_z \parallel  [100])$ or in the sample plane $(B^1_x \parallel  [011])$.
\section{Warm-up spectroscopy method}

\begin{figure}
	\includegraphics[width=3.3in]{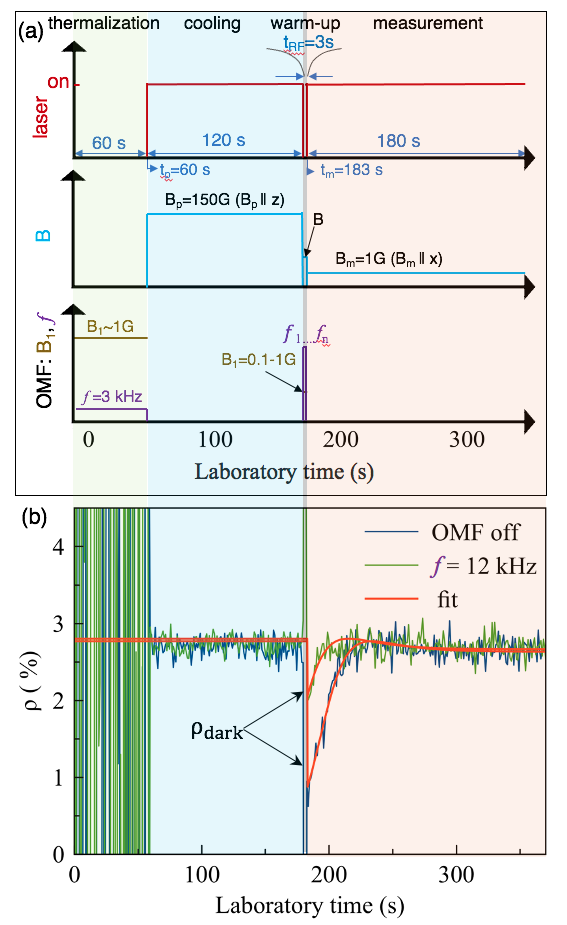}

	\caption{(a) Timing diagram of the experimental cycle including thermalization with the lattice ($60$~s), optical cooling ($120$~s), adiabatic demagnetization ($20$~ms), application of the OMF (warm-up stage, $3$~s) and measurement. (b) Evolution of the PL polarization with (green line) and without (blue line) OMF during the cycle. Red curves are fits of the data to Eq.~\eqref{GrindEQ__3_}. The difference between  $\rho_{\mathrm{dark}}$ values for blue and green curves characterizes the effect of OMF heating on the NSS.}
			\label{fig:timing}
\end{figure}

The experimental protocol that we implement for warm-up spectroscopy is illustrated in Fig.~\ref{fig:timing}. 
Each point of the warm-up spectrum corresponding to the OMF frequency $f$ is measured in a sequence of four stages described below:

{\it 1. Thermalization. } We start from erasing any NSS polarization and setting up the  NSS temperature $\theta_N^{\infty}$ such that $1/\theta_N^{\infty}\approx 0$. To do so we wait "in the dark" (laser off) at $B=0$ during $60$~s in presence of the erasing OMF at the frequency $f=3$~kHz. 

{\it 2. Cooling. } The cooling stage starts at $t=t_p=60$~s. It consists of NSS cooling via optical pumping and adiabatic demagnetization. During $120$~s of optical pumping the longitudinal field  $B_p=150$~G is applied, and the circularly polarized laser beam is switched on. Under optical pumping, nuclear magnetization builds up due to hyperfine interaction between electron spins localized at the donors and the underlying nuclei. This magnetization spreads out within the crystal by spin diffusion \cite{PhysRevB.25.4444}. Typical diffusion distance can be estimated as $l_d \approx 50$~nm 
\footnote{
Assuming $D=10^{13}$~cm$^2$/s \cite{PhysRevB.25.4444}, pumping time $t_p=60$~s and $l_d=2\sqrt{D t_p}$
}.
The time dependence of the Overhauser field in the external static magnetic field $B$ is controlled by the time dependence of the nuclear spin temperature. It is given by the expression \cite{OO1}: 
\begin{equation} \label{GrindEQ__2_} 
B_N(t)=B \frac{b_N h{\overline{\gamma}}_N}{k_B}
\frac{I(I+1)}{3} \beta_N(t), 
\end{equation} 
where 
$\beta_N\left(t\right)={1}/{\theta_N\left(t\right)}$ is the inverse nuclear spin temperature that increases under optical pumping starting from ${1}/{\Theta_N^{\infty}}$;  $b_N=5.3$~T is the Overhauser field at saturation of magnetization; ${\overline\gamma}_N$ is the average gyromagnetic ratio of the NSS of GaAs; \textit{h} and $k_B$ are Planck and Boltzmann constants, respectively. After $120$~s optical pumping the laser is switched off and magnetic field is adiabatically (in $20$~ms) swept down to zero.  This reduces the nuclear spin temperature $\theta_N$ reached after optical cooling down to  
$\theta_N^{ad}$, according to Eq.~(\ref{GrindEQ__1_}).

 Note that deep NSS cooling is not mandatory but quite useful to achieve high magnetic susceptibility of the NSS and increase its response to OMF \cite{Abragam}. Indeed, according to Eq.~\eqref{GrindEQ__2_}, nuclear magnetization in the external field is proportional to the inverse spin temperature $\beta_N$. Therefore, the measured Overhauser field  can be significant even in a very low external magnetic field, provided that NSS is cold. 
 In this work after optical pumping and adiabatic demagnetisation to $B=0$, conducted within the protocol described above, the Overhauser field $B_N=10$~G builds up in the measurement field $B_m=1$~G. This corresponds to nuclear spin temperatures $\theta^{ad}_N \approx 300$~$\muup$K at $B=0$, see Eqs.~(\ref{GrindEQ__1_})-(\ref{GrindEQ__2_})
 \footnote{In bulk GaAs crystals, the Overhauser field under optical pumping can reach several Tesla. We deliberately use weak measurement fields and moderate nuclear spin temperatures to match the Overhauser field at the measurement stage with the width of the electron Hanle effect in our sample}.

{\it 3. Warm-up. } The heating of the NSS by the OMF is the shortest, but the most essential stage of this protocol. During this stage the laser remains switched off, so that there are no out-of-equilibrium carriers in the sample. The static field is set at a value of interest, $B$ (either $B_\mathrm{x}$ or $B_\mathrm{z}$), and the OMF $B_1=0.1 - 1$~G is applied during $t_\mathrm{RF}=3$~s at a fixed frequency \textit{f} ranging from $0.1$ to $100$~kHz. The amplitude of the OMF, $B_1$, is chosen for each static magnetic field in such a way, that it erases about $70$~\% of nuclear polarization at the frequency of absorption maximum. Then each spectrum is normalized to the square of $B_1$ \footnote{The intensity of the incident  field is proportional to $B_1^2$}. We expect stronger absorption  when the frequency of the OMF  matches one of the nuclear spin transition frequencies, so that  the NSS temperature increases faster than under non-resonant excitation. 

{\it 4. Measurement. } The goal of the last stage is to determine the OMF-induced enhancement of the NSS warm-up rate. To do so, at $t=t_\mathrm{m}=183$~s the pump beam and the static in-plane magnetic field $B_\mathrm{m}=1$~G are turned on. This results in the polarized PL that is carefully recorded during at least $180$~s, as illustrated in Fig.~\ref{fig:timing}~(b). 
The PL polarization $\rho \left(t\right)$ in the transverse magnetic field is known to decrease with the field strength, following the Lorenzian dependence (the Hanle effect) \cite{OO1}. The Overhauser field of the cooled nuclear spins builds up in our probe magnetic field, being parallel or antiparallel to it, depending on the sign of spin temperature. The total field acting on the electron spins is therefore equal to $B_\mathrm{tot}(t)=B_\mathrm{m} \pm B_N(t)$. The Overhauser field $B_N$ decreases with the characteristic nuclear spin relaxation time $T_1$. The evolution of the PL polarization follows the Hanle curve in the time-dependent total field:
\begin{equation} \label{GrindEQ__3_} 
\rho \left(t\right)=\frac{ \rho_0  B^2_{1/2}}
{{B^2_{1/2}+
\left(B_\mathrm{m}+b+(B_N(t_\mathrm{m})-b)\times \exp(-\frac{\left(t-t_\mathrm{m}\right)}{T_1})\right)}^2}, 
\end{equation} 
This expression accounts for a weak polarization of nuclear spins during the measurement stage, which occurs due to the nuclear spin cooling in the Knight field of electrons \footnote{The consistent theory of the Hanle effect in presence of dynamic polarization of nuclear spins (including that in the Knight field of photoexcited electrons) is given in \cite{OO1}. It says that the magnetization of cooled nuclei in the Knight field does not affect the Hanle curve, since the resulted Overhauser field is parallel to the mean electron spin. As a consequence, the only contribution of the Knight field comes from additional cooling of the NSS, proportional to ${S_0}{S_z}$, where $S_0$ is the initial mean spin of photoexcited electrons at birth, and $S_z$ is their current spin projection on the excitation axis. Our experiments were designed so that the variation of the nuclear spin temperature due to cooling in the Knight field was much smaller than that due to preliminary optical cooling. It is only important at the very end of the spin-lattice relaxation of the NSS, as seen from Fig.~\ref{fig:timing}~(b). At this stage, the relative variation of $S_z$ (proportional to the PL polarization $\rho \left(t\right)$) is small, which justifies using a linearized rate equation of dynamic polarization, leading to the single-exponential time dependence of the Overhauser field.}.
This effect results in building up a nuclear field $b\approx 0.2$~G. As seen from Fig.~\ref{fig:timing}~(b), the PL polarization $\rho\left(t\right)$ initially increases, starting from the value $\rho_{dark}$, and eventually approaches the stationary PL polarization defined by optical pumping in the probe field. By fitting Eq.~\eqref{GrindEQ__3_} to the experimental data we determine the Overhauser field immediately after the heating stage,  $B_N(t_\mathrm{m})\equiv B_N(f)$. In the experiments presented below, NSS is cooled down to positive temperatures, so that the Overhauser field is antiparallel to the external field, but we have checked that the results do not depend on the spin temperature sign. 


The rate of the NSS heating  ${1}/T_\mathrm{RF}(f)$ induced by the OMF at frequency $f$, can be expressed in terms of the Overhauser field before ($B_{N0}$) and after ($B_N(f)$) heating  \cite{OO1}:
\begin{equation} \label{GrindEQ__4_} 
B_N(f)=B_{N0}\ \mathrm{exp}\left({-t_\mathrm{RF}\left(\frac{1}{T_1}+\frac{1}{T_\mathrm{RF}(f)}\right)}\right).                                 \end{equation} 
Measuring the calibration curve in the absence of the OMF (blue line in Fig.~\ref{fig:timing}~(b)) allows us to simplify Eq.~(\ref{GrindEQ__4_}) and to use the following expression for the warm-up rate:
\begin{equation} \label{GrindEQ__5_} 
\frac{1}{T_\mathrm{RF}}(f)=\frac{1}{t_\mathrm{RF}}{\mathrm{ln}\left(\frac{B^\prime_N}{B_N(f)}\right)},                          \end{equation} 
where $B^\prime_N$ is the Overhauser field at $t=t_\mathrm{m}$ in the absence of the OMF. 

The four-stage procedure described above allows us to determine the warm-up rate at a  given value of the OMF frequency. By repeating the entire  protocol for each frequency from $0.1$ to $100$~kHz with the step of $1$~kHz we obtain the warm-up spectrum at a given static field and orientation of OMF. Typical time for one spectrum  acquisition is of order of $10$ hours. 
\section{Experimental results }
\begin{figure}
\includegraphics[width=3.1in]{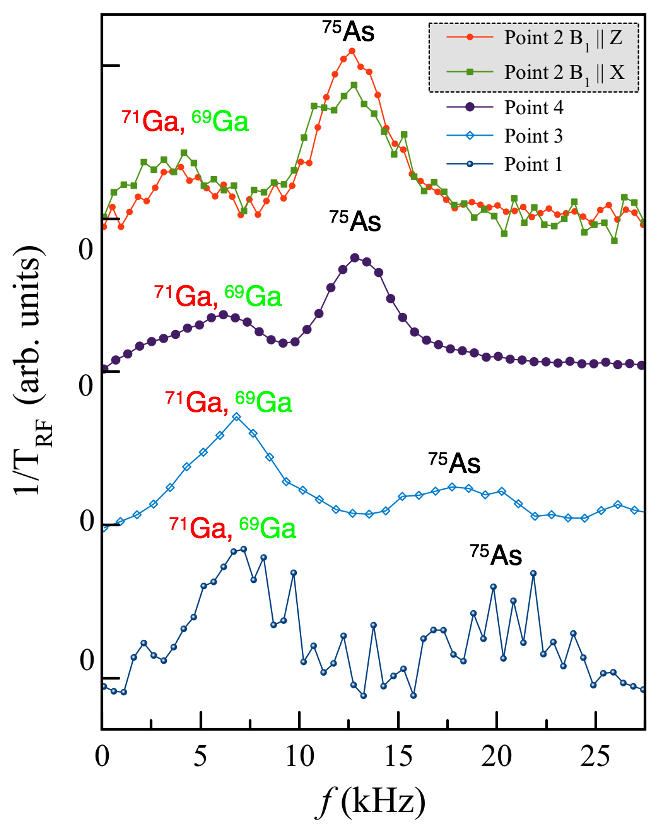}
\caption{Warm-up spectra measured at zero static magnetic field in four different points on the sample surface. Point 2 is the one studied throughout the paper. We identify $^{75}$As resonance at higher frequency and $^{71}$Ga/$^{69}$Ga  resonances merged together at lower frequency.  The significant difference between spectra suggests that quadrupole splittings of all isotopes are inhomogeneous in the sample plane.}  
	\label{fig:B=0}
\end{figure}
\begin{figure}
\includegraphics[width=3.2in]{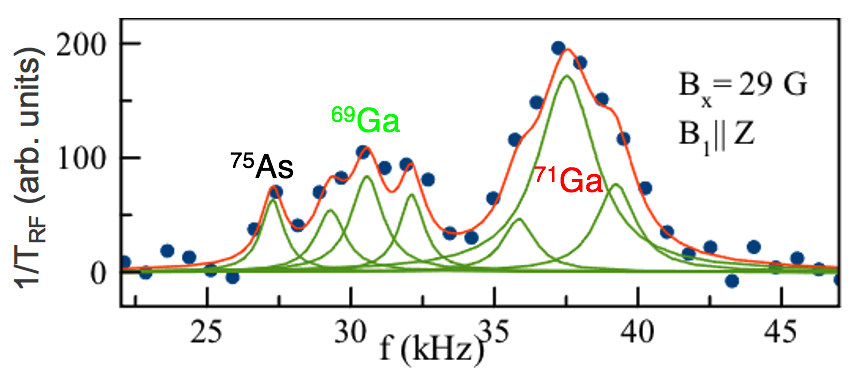}
\caption{Warm-up spectra (symbols) measured in orthogonal configuration $B \parallel x$, $B_1 \parallel z$ at point 2 on the sample surface (\textit{cf} Fig.~\ref{fig:B=0}). For both Ga isotopes quadrupole satellites can be distinguished. Green solid lines are Gaussian contributions and red line is their sum obtained via least squares fitting of the spectra.}
	\label{fig:fits}
\end{figure}

Warm-up spectra measured at zero static magnetic field in four different points on the sample surface are shown in Fig.~\ref{fig:B=0}. According to ideas illustrated in  Fig.~\ref{fig:fig1},  we expect to see for each isotope one resonance corresponding to the transition  between two quadrupole-split doublets: ($1/2$, $3/2$) and ($-1/2$, $-3/2$). We attribute the lower frequency peak to $^{71}$Ga and $^{69}$Ga resonances that are merged together since the values of their quadrupole moments $Q$ are quite close, and the higher frequency peak to $^{75}$As that has the highest quadrupole moment, see Table 1.

The significant difference between the  spectra  measured at different points (up to $50\%$  variation of the resonant frequency) indicates that the deformation and electric field that could contribute to the quadrupole interaction are strongly inhomogeneous in the sample plane. This is not surprising  as far as we deal with very small strain \cite{Yamada1998l}.  We have chosen Point 2 for the detailed analysis as a function of magnetic field presented below.

Under magnetic field $B$ the spin states for each isotope are expected to split in four, and the dependence of the nuclear spin resonances is \textit{a priori} determined by the ratio of the quadrupole and Zeeman energies. Thus, for each of three isotopes four different transitions are expected at $B \perp  B_1$ and two when $B \parallel B_1$  (at least at $B \leq B_L$), see arrows in Fig.~\ref{fig:fig1}.  

A typical warm-up spectrum measured at $B \parallel x$, $B_1 \parallel  z$  is shown in Fig.~\ref{fig:fits}. Despite relatively low spectral resolution as compared to the linewidth, one can guess multiple contributions to the spectrum that one can  identify as (i) $-1/2 \leftrightarrow 1/2$ transitions for all isotopes at frequencies given by corresponding Zeeman splittings (red solid arrows in Fig.~\ref{fig:fig1}) and  (ii) a pair of satellite lines  for each of Ga isotopes. These quadrupole-induced satellites result from two other $\pm 1$ spin transitions as illustrated in  Fig.~\ref{fig:fig1} by red dashed and dotted arrows. $^{75}$As satellites seem to be weaker and the higher frequency satellite line is certainly masked by the close-lying $^{69}$Ga.


%
\begin{figure*}
\includegraphics[width=6.5in]{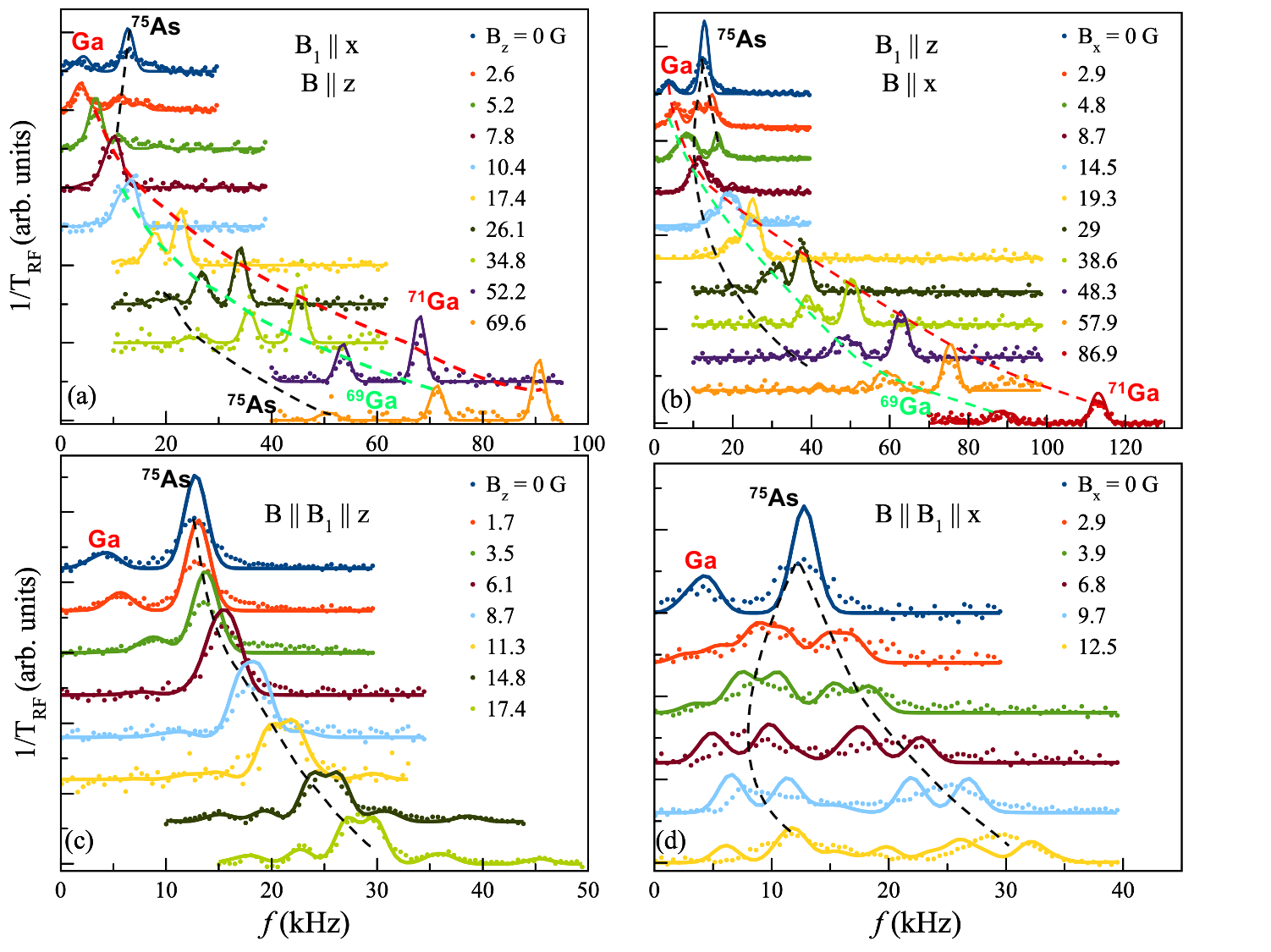}
\caption{Waterfall-arranged warm-up spectra (points) measured at different values of $B \parallel B_z$ (a, c) and $B \parallel B_x$ (b, d). OMF ($B_1$) is oriented either along \textit{z}-axis (b, c) or along \textit{x}-axis (a, d). Each spectrum is normalized to its integrated intensity, shown in Fig.~\ref{fig:TotAbs}. Solid lines are the results of least squares fit of the model based on Eq.~\eqref{GrindEQ__7_} to these the spectra, assuming fixed transition linewidth $1.5$~kHz. Dashed lines on top of the plots are guide for the eye, they sketch the field dependence of the various spin transitions. }
	\label{fig:allfits}
\end{figure*}

\begin{figure}
\includegraphics[width=3.0in]{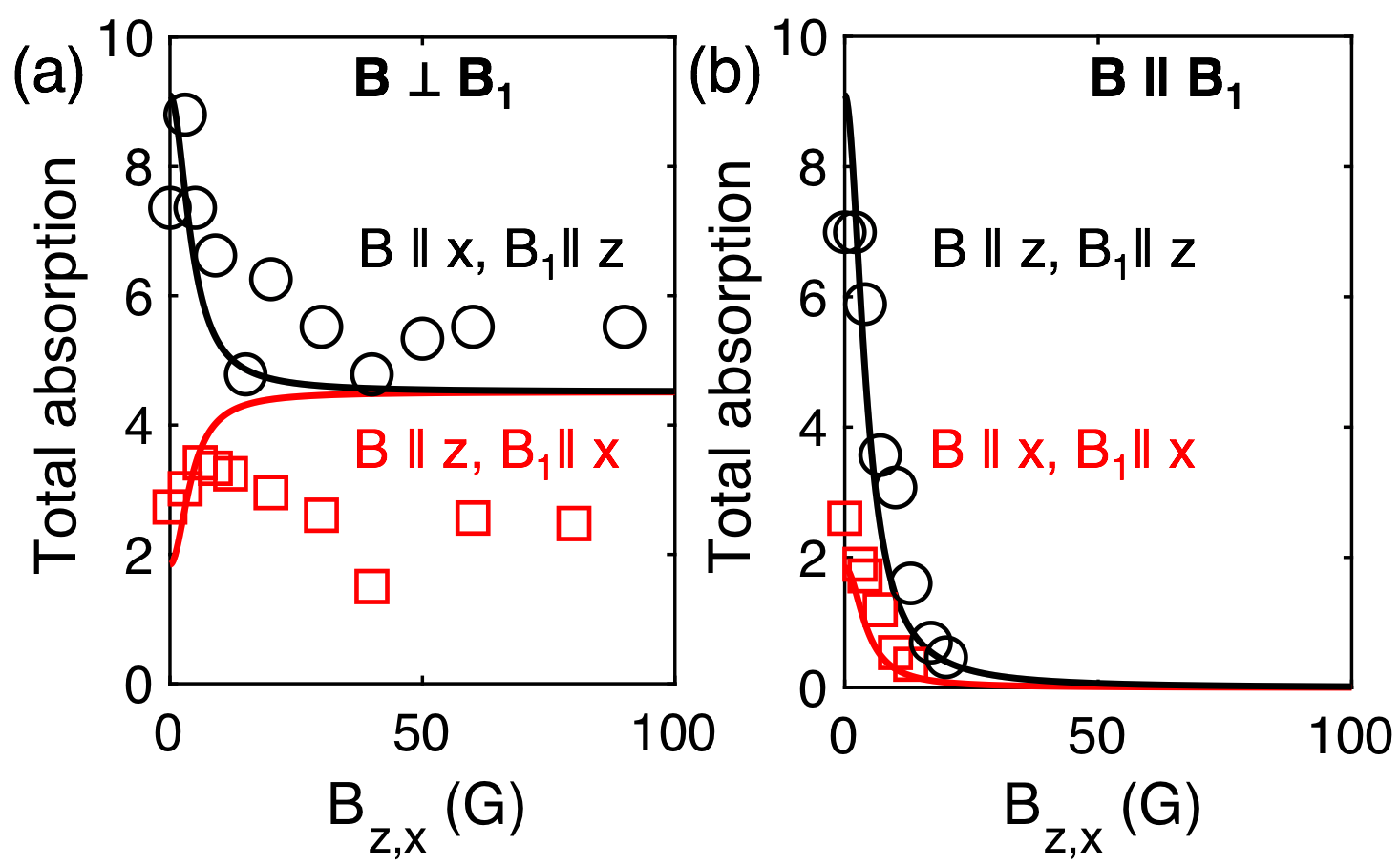}
\caption{Integrated intensity of the warm-up spectra (circles) as a function of the static magnetic field in longitudinal (a) and transverse (b) experimental configurations. Solid lines show the same quantities obtained from the fit.}
	\label{fig:TotAbs}
\end{figure}

Fig.~\ref{fig:allfits} gathers waterfall-arranged warm-up spectra (points) measured at different relative orientations of static field and OMF, either perpendicular (top panels) or parallel (bottom panels) as well as their orientation with respect to the growth axis. For convenience of comparison between the spectra, each spectrum is normalized to its integrated intensity. {The corresponding values of the integrated intensity are shown in Fig.~\ref{fig:TotAbs}}. 

In orthogonal geometries ($B \perp B_1$, Fig.~\ref{fig:allfits}~(a-b)),  $^{75}$As peak rapidly fades out  ({\it cf} black dashed line), and is hardly distinguishable at highest fields. By contrast  Ga  peak splits and its intensity increases. At high fields it  evolves into Zeeman frequency for $^{71}$Ga (red dashed line) and $^{69}$Ga (green dashed line) and two equidistant pairs of poorly resolved satellites. These satellites are better distinguished for $^{69}$Ga in Fig.~\ref{fig:allfits}~(b).  

 In parallel geometries ($B \parallel B_1$, Fig.~\ref{fig:allfits}~(c-d)) the RF absorption remains measurable up to magnetic fields of order of local field $B_L$  due to the presence of the quadrupole effects.
One can see that Ga contributions do not grow up with magnetic field as in orthogonal geometries.  The   behaviour of  $^{75}$As depends on the orientation of the fields with respect to the growth axis.  
If both fields lie in the plane of the sample, $^{75}$As peak splits into two broad peaks. If the fields are perpendicular to the growth axis, the $^{75}$As peak does not split with the field, but its frequency grows quadratically with magnetic field. 

We show below that this dense forest of spectral features can be unraveled and understood as a result of the interplay between quadrupole and Zeeman interaction in the system with three different isotopes.

\section{Theory }

The quadrupole interaction of a nucleus with spin $I$ and quadrupole moment $Q$ with EFGs is described by the following Hamiltonian \cite{Abragam}:
\begin{equation}
\label{GrindEQ__6_} 
 \hat{H}_Q=
 \frac{eQ}{6I(2I-1)}\sum_{j,k} V_{jk}
 \left(
 \frac{3}{2} 
 \left(\hat{I}_j\hat{I}_k+\hat{I}_k \hat{I}_j\right)
 -\delta_{jk}\hat{I}^2 
 \right), 
\end{equation} 
where $V_{jk}$ is the EFG tensor at the nucleus location, $e$ is electron charge. In GaAs all isotopes are characterized by $I=3/2$.
In unperturbed cubic crystals, the elements of $V_{jk}$ are {equal to} zero. {But} EFG may arise due to lattice deformation and, because of absence of inversion symmetry in the zinc-blende lattice, due to homogeneous electric fields.  Our sample is a relatively thin platelet with one of the axes of zinc-blende lattice directed along the \textit{z-}axis normal to the sample plane. In this case, one can expect shear strains in \textit{xz} and \textit{yz} planes to be zero. Electric fields, if present, created by the surface charge or doping gradient, are always directed along \textit{z}-axis \cite{PhysRevB.76.245301,King2012}.  Under these assumptions, the Hamiltonian of the NSS in the presence of quadrupole interaction reads:
\begin{equation} \label{GrindEQ__7_} 
 \hat{H}_{Q}^i=\frac{E_{QZ}^i}{2}
 \left(\hat{I}_z^2 -\frac{I(I+1)}{3}\right)+
 \frac{E_{QR}^i}{4\sqrt{3}}
 \left(\hat{I}_+^2 + \hat{I}_-^2\right)+
 \mathrm{i}\frac{E_{QI}^i}{4\sqrt{3}}
 \left(\hat{I}_+^2 - \hat{I}_-^2\right),
\end{equation} 
where  $i=$ ${}^{71}$Ga, ${}^{69}$Ga and ${}^{75}$As and $\mathrm{i}$ is the unit imaginary number. It is characterized by 9 parameters, 3 for each isotope.
The energy parameter $E_{QZ}$ {determines} the uniaxial deformation along \textit{z}-axis:
\begin{equation} \label{GrindEQ__8_} 
 E^i_{QZ} =\frac{3eQ^iS^i_{11}}{I(I+1)}\left({\varepsilon }_{zz}-\frac{{\varepsilon }_{xx}+{\varepsilon }_{yy}+{\varepsilon }_{zz}}{3}\right),                                              
\end{equation} 
where  $S_{11}$ is a diagonal element of the fourth rank gradient-elastic tensor, ${\varepsilon }_{zz},{\varepsilon }_{yy}\ $and ${\varepsilon }_{xx}$ are diagonal components of the second rank elastic strain tensor. 
The energy parameter $E_{QR}$ accounts for the uniaxial deformation along \textit{x} and \textit{y}-axes:
\begin{equation} \label{GrindEQ__9_} 
E^i_{QR}=\frac{3\sqrt{3}eQ^iS^i_{11}}{I\left(I+1\right)}\left({\varepsilon }_{xx}-{\varepsilon }_{yy}\right).
\end{equation} 
The energy parameter $E_{QI}$ is related to the shear strain in the $xy$-plane and/or to a built-in electric field $F$ along the growth axis:
\begin{equation} \label{GrindEQ__10_} 
E^i_{QI}=\frac{3\sqrt{3}eQ^iS^i_{44}}{2I\left(I+1\right)}{\varepsilon }_{xy}+\frac{3\sqrt{3}eQ^iR^i_{14}}{2I\left(I+1\right)}F. \end{equation} 
Here  $S_{44}$ is an off-diagonal component of the gradient-elastic tensor, $R_{14}$ is the tensor component relating EFG with homogeneous electric field $F$ oriented along $z$-axis \cite{PhysRevB.20.4406}, and ${\varepsilon }_{xy}$ is a off-diagonal component of the elastic strain tensor. Elastic strain and electric field in these equations are common for all isotopes, while quadrupole moments and gradient-elastic tensors are different (see Table 1). 
Note, that the values of the gradient-elastic tensor components $S_{11}$ and $S_{44}$ has been recently determined by nuclear magnetic resonance spectroscopy of GaAs/AlGaAs quantum dot structures \cite{PhysRevB.97.235311,Griffiths2019} and our experiments provide an opportunity to check them in another kind of experiments.

We can further assume that the stress experienced by {the} sample is a combination of a compressive pressure $p_1$ in \textit{xy} plane applied at some angle $\varsigma $ with respect to \textit{x}-axis and a decompressive pressure $p_2$ in the orthogonal direction. In this case all components of the strain tensor ${\varepsilon }_{jk}\ $can be calculated from $p_1$, $p_2$ and $\varsigma$ using the stiffness tensor $C_{iklm}$ and cubic symmetry of the crystal:
\begin{equation} \label{GrindEQ__11_} 
{\varepsilon }_{xx}=\frac{\left(C_{11}+2C_{12}\right)\left(p_1{{\mathrm{cos}}^2 \varsigma \ }+p_2{{\mathrm{sin}}^2 \varsigma \ }\right)-C_{12}\left(p_1+p_2\right)}{\left(C_{11}-C_{12}\right)\left(C_{11}+2C_{12}\right)},  
\end{equation} 
\begin{equation} \label{GrindEQ__12_} 
{\varepsilon }_{yy}=\frac{\left(C_{11}+2C_{12}\right)\left(p_1{{\mathrm{sin}}^2 \varsigma \ }+p_2{{\mathrm{cos}}^2 \varsigma \ }\right)-C_{12}(p_1+p_2)}{\left(C_{11}-C_{12}\right)\left(C_{11}+2C_{12}\right)},                                           
\end{equation} 
\begin{equation} \label{GrindEQ__13_} 
\ {\varepsilon }_{zz}=-\frac{C_{12}(p_1+p_2)}{\left(C_{11}-C_{12}\right)\left(C_{11}+2C_{12}\right)},                           \end{equation} 
\begin{equation} \label{GrindEQ__14_} 
{\varepsilon }_{xy}={\varepsilon }_{yx}=\left(p_1-p_2\right)\frac{{\mathrm{cos} \varsigma \ }{\mathrm{sin} \varsigma \ }}{C_{44}}\ . 
\end{equation} 
Remaining ${\varepsilon}_{lm}$ components are equal to zero. Thus, we can rewrite Eqs.~\eqref{GrindEQ__8_} - \eqref{GrindEQ__10_} in terms of pressure parameters $p_1$ and $p_2$ and the angle $\varsigma$:
\begin{equation} \label{GrindEQ__15_} 
E_{QZ}^i=\frac{eQ^iS_{11}^i}{I\left(I+1\right)}\frac{1}{C_{11}-C_{12}}\left(p_1+p_2\right), 
\end{equation} 
\begin{equation} \label{GrindEQ__16_} 
E_{QR}^i=\frac{3\sqrt{3}eQ^iS_{11}^i}{I\left(I+1\right)}\frac{{\mathrm{cos} \left(2\varsigma \right)\ }}{C_{11}-C_{12}}\left(p_1-p_2\right),\  
\end{equation} 
\begin{equation} \label{GrindEQ__17_} 
E_{QI}^i=\frac{3\sqrt{3}eQ^iS_{44}^i}{4I\left(I+1\right)}\frac{{\mathrm{sin} \left(2\varsigma \right)\ }}{C_{44}}\left(p_1-p_2\right)+\frac{3\sqrt{3}eQ^iR_{14}^i}{2I\left(I+1\right)}F. 
\end{equation} 

\begin{table}
\begin{ruledtabular}
\begin{tabular}{|c c c c|}
\multicolumn{4}{|c|}{GaAs parameters} \\ 
\hline 
\hline 
$C_{11}\times {10}^{10}($ ${\mathrm{N}}/{{\mathrm{m}}^{\mathrm{2}}})$ \cite{SS73} &   &    12 &   \\ 
\hline 
$C_{12}\times {10}^{10}$ (${\mathrm{N}}/{{\mathrm{m}}^{\mathrm{2}}})$  \cite{SS73}&   &   5.4 &   \\ 
\hline 
$C_{44}\times {10}^{10}$ (${\mathrm{N}}/{{\mathrm{m}}^{\mathrm{2}}})$  \cite{SS73} &  &    6.2 &  \\ 
\hline 
\hline
Isotope Parameters & ${}^{71}$Ga  &   ${}^{69}$Ga &   ${}^{75}$As \\ 
\hline
\hline 
{$A$}& {$0.2$}& {$0.3$}& {$0.5$ }\\
\hline 
$S_{11}\times {10}^{-21}$ ( ${\mathrm{V}}/{{\mathrm{m}}^{\mathrm{2}}}$ ) \cite{PhysRevB.97.235311} &  -22 &  -22  &  24.2 \\ 
\hline 
$S_{44}\times {10}^{-21}$ ( ${\mathrm{V}}/{{\mathrm{m}}^{\mathrm{2}}}$ ) \cite{PhysRevB.97.235311} & { 9} &  9 &  {\bf 48} \\ 
\hline 
 $S_{44}\times {10}^{-21}$ ( ${\mathrm{V}}/{{\mathrm{m}}^{\mathrm{2}}}$ )  [this work]  & { 9} &  9 &  {\bf 17$\pm 3$} \\ 
\hline
$R_{14}\mathrm{\times }{10}^{12}\ ({\mathrm{m}}^{\mathrm{-}\mathrm{1}})$  \cite{PhysRev.129.1965} &  2.1&  2 &  1.9\\ 
\hline 
$Q\times {10}^{30}$ (m${}^{2}$) \cite{doi:10.1080/00268970802018367} &  10.7 & 17.1 &  31.4 \\ 
\hline 
$\gamma_N $ (kHz/G) \cite{doi:10.1080/00268970802018367} &  0.82 &  0.64 &  0.45 \\ \hline 
$E_{QZ}$/h (kHz) & 0.45 & 0.73 &   -1.5 \\ 
\hline 
$E_{QR}$/h (kHz) &  -2 &   -3.3 &  6.5 \\ 
\hline 
$E_{QI}$/h (kHz) &   -2 &   -3.1 &    -11 \\
\end{tabular}
\end{ruledtabular}
\caption{Values of the {abundance}, relevant gradient-elastic, electric field and stiffness tensors elements, quadrupole moments and gyromagnetic ratios for three GaAs isotopes.
Last three lines show the quadrupole frequencies resulting of the fitting procedure. }
\end{table}

The total energy of the NSS in the external magnetic fields is given by the sum of the quadrupole and Zeeman Hamiltonians for each isotope:
\begin{equation} \label{GrindEQ__18_} 
\hat{H}^i=\hat{H}_{Q}^i+h\gamma_N^i(\hat{I}\cdot B),            \end{equation} 
were $\gamma_N^i$ are nuclear gyromagnetic ratios. The energy levels of each isotope are given by the eigenvalues of ${\hat{H}}^i$. 
The dipole-dipole interaction which is known to broaden the NMR transitions, but does not lead to any shifts of their energies, is not included here\cite{Abragam}.
The eigenfunctions $\mid\Psi_m\rangle$ of the Hamiltonian ${\hat{H}}^i$ are superpositions of the states with angular momentum  $\mid\pm 1/2\rangle$ and 
$\mid \pm 3/2 \rangle$ that depend on the orientation of the pressure axes with respect to the crystallographic directions.
They can be found by numerical diagonalization of the Hamiltonian. The OMF induces spin transitions between {a pair of states} if its frequency matches the energy difference between {their} energy levels. Since all the states are mixed, the OMF can induce transitions between any pair of spin energy levels of a given isotope. The probability of transition $P_{kl}^i$  between energy levels $E_k$ and $E_l$ of $i$-th isotope is given by  $P_{kl}^i \propto M^2_{kl,i}$, where $M_{kl}=\langle \Psi _k\left|H_{OMF}\right|\Psi_l\rangle$ is the matrix element of the Hamiltonian describing Zeeman interaction of the nuclear spins with the OMF. Thus, the warm-up rate associated with each transition   reads
\begin{equation} 
\label{eq:wurate_theor} 
\frac{1}{T_{RF}}\bigg|_{i, kl}=
\frac{A^i P_{kl}^i  \left| E^i_k-E_l^i \right|^2}
{\sum_i{A^i}\sum_{j=1}^{6}\left| E_j\right|^2},
\end{equation} 
{where $A^i$ stands for the isotope abundance}. Note, that the denominator in Eq.~\eqref{eq:wurate_theor} represent the heat capacity of the NSS per one nucleus.

\section{Experiment versus theory: discussion}

\begin{figure*}
\includegraphics[width=6.80in]{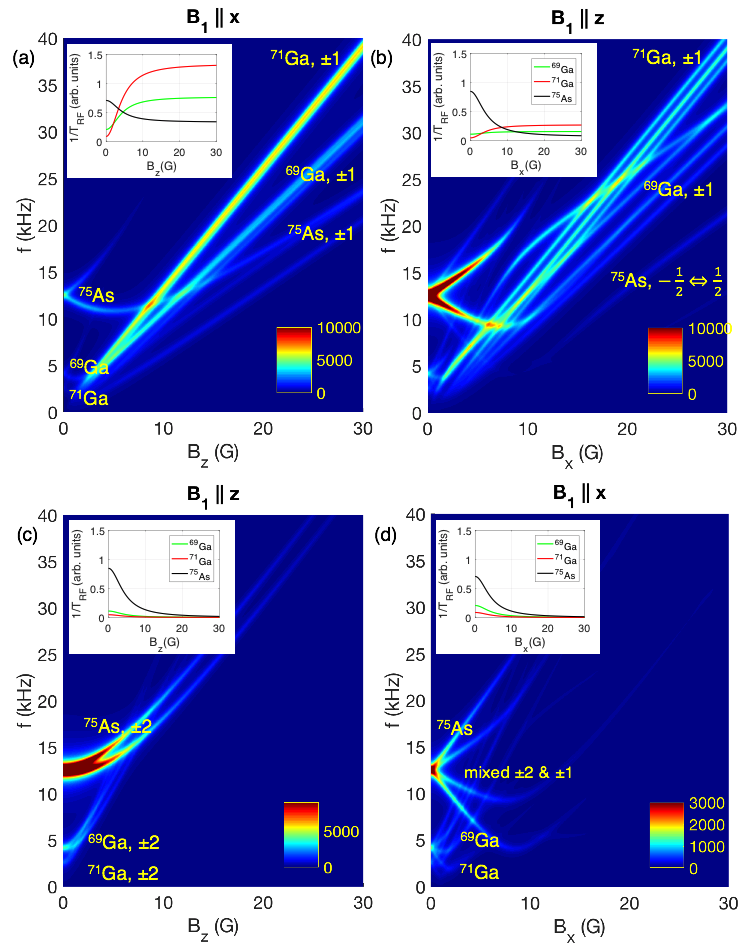}
\caption{Color-encoded warm-up spectra obtained by fitting of the experimental spectra shown in Fig.~\ref{fig:allfits} to the model defined by Eqs.~\eqref{GrindEQ__7_} and \eqref{GrindEQ__18_}  at different values of $B \parallel B_z$ (a, c) and $B \parallel B_x$ (b, d). OMF is oriented either along \textit{z}-axis (b, c) or along \textit{x}-axis (a, d). The corresponding  parameters are given in Table 1.}
	\label{fig:3D}
\end{figure*} 
To fit the above model to the experimental data (the ensemble of 35 spectra shown in Fig.~\ref{fig:allfits}) we proceed as follows.
First, using quadrupole energies  as parameters, we calculate the transition frequencies and transition probabilities for all the values and orientations of magnetic fields and for all isotopes. To transform these values into theoretical spectra, we convolute them with Gaussians using an empirically defined half-width $\delta f=\ $1.5 kHz. It is determined by the dipole-dipole interactions within NSS. Then, we iterate over fitting parameters  $E_{QZ}$, $E_{QR}$ and $E_{QI}$, in order to minimize the total mean square difference between experimental and theoretical spectra. The best fit values are given in Table~1.

Theoretical spectra with these parameters are shown in Fig.~\ref{fig:allfits} by solid lines, and the agreement is surprisingly good. Note, however, that to fit relative intensities of some peaks in the spectra at different fields we introduced a minor disorientation of the OMF with the crystal axes, $3^\circ$ for the in-plane field, and $7^\circ$ for the field along \textit{z}-axis.  This misalignment is due to uncontrollable shifts of the small size RF coils that cannot be repositioned any more when the sample is cooled down.

In Fig.~\ref{fig:TotAbs}(a), (b) we compare the calculated integrated intensities with the experimentally measured total RF absorption.  In this calculation we assume that spins of different isotopes are thermalized, so that the heat capacity is determined by the entire NSS. Under this assumption the magnetic field dependence of the total warm-up rate is well reproduced, suggesting that the thermodynamic description of the quadrupole-split NSS and its heating by OMF is, indeed, justified.

Fig.~\ref{fig:3D} represents color-encoded theoretical spectra as a function of the magnetic field in four relevant experimental geometries. The contributions of the three isotopes and the characteristic spin transitions are identified. Insets show relative total absorption of the three isotopes.
%

We can now identify better the main spin transitions that appear in experiments. 
 Let us first consider $^{75}$As. Its contribution to the total absorption always decreases with magnetic field because at $B=0$ it is  determined by the quadrupole moment, that is highest for $^{75}$As, while at high field it is essentially proportional to the square of the gyromagnetic ratio, that is lowest for  $^{75}$As compared to other isotopes (insets in Fig.~\ref{fig:3D}). This is indeed observed in Fig.~\ref{fig:allfits}. Another salient feature is the  behaviour of the $^{75}$As resonance at low fields in parallel geometries (Figs.~\ref{fig:allfits}~(c, d), \ref{fig:3D}~(c, d)). It is expected to be dominated by $\pm 2$ spin transitions. However, at low in-plane magnetic  field  the four spin states are strongly mixed due to the large in-plane quadrupole energies $E_{QR}$ and $E_{QI}$, see also  Fig.~\ref{fig:fig1}, where the parameters are chosen to be the same as the result of the fitting procedure for $^{75}$As. This mixing results in a large in-plane magnetic  field-induced  splitting of $^{75}$As absorption line, as highlighted by a black dashed line on top of Fig.~\ref{fig:allfits}~(d). This mixing is not important at $B \parallel B_z$, but the large in-plane quadrupole energies lead to quadratic frequency shift, highlighted in Fig.~\ref{fig:allfits}~(c) by a black dashed line. 
 
 The two Ga isotopes are best visible in perpendicular geometries for the reasons already mentioned above: both have small quadrupole moment but high gyromagnetic ratio as compared to $^{75}$As. 
 At high fields three $\pm 1$ spin transitions for each Ga isotope are expected to dominate the spectrum: central transition (CT) $-1/2 \leftrightarrow 1/2$ and two satellite transitions (ST):   $-3/2 \leftrightarrow -1/2$, $1/2 \leftrightarrow 3/2$, {\it cf} also Fig.~\ref{fig:fig1}.  The frequency of the CT nicely follows Zeeman splitting dependence on the magnetic field, see green  and  red dashed lines for $^{69}$Ga and $^{71}$Ga, respectively. The isotope-dependent frequency shift between CT and ST depends on the orientation of the static field. This is clearly seen in Figs.~\ref{fig:3D}~(a-b). One can show that for $B_z$ the frequency shift  is given by $\Delta f_{QZ}^i= \lvert E_{QZ}^i \rvert$  and for $B_x$ by 
 \begin{equation} 
\label{eq:sat} 
 \Delta f_{QX}^i= \frac{1}{h} \left | \frac{E_{QZ}}{2} -\frac{\sqrt{3}}{2} \sqrt{E_{QR}^2+E_{QI}^2} \mathrm{cos}2 (\zeta-\phi) )\right |,
 \end{equation}
 where $\mathrm{tg}(2\zeta)=E_{QI}/E_{QR}$, and $\phi$ is the angle between $B_x$ and $[100]$ axis (we assume in the following that in our experiments $\phi=\pi/4$ corresponds to $B_x \parallel [110]$ although this choice is somehow arbitrary: the orientation of the crystallographic axes is known to the $\pi/2$  factor only). 
 Although the experimentally observed CT-ST frequency shifts are small compared to the linewidth (see also Fig.~\ref{fig:fits}), they are clearly smaller in the presence of $B_x$ as compared to $B_z$, $\Delta f_{QX}^i< \Delta f_{QZ}^i$. This also indicates presence of EFG components in the plane of the sample.

The fitting procedure points out that all the quadrupole energies $E_{QZ}$, $E_{QR}$ and $E_{QI}$ defined by Eqs.~\eqref{GrindEQ__8_}-\eqref{GrindEQ__10_} are different from zero. Thus, there should be at least two sources of the quadrupole interaction in our sample: tension/compression and either shear strain, or built-in electric field.
%
Assuming the NSS parameters known from the recent work by Griffiths {\it et al} \cite{Griffiths2019} as summarized in Table 1, and in particular the ratio $S_{11}/S_{44}$ for different isotopes, this  mechanical deformation is not sufficient, and both strain and an electric field $F$ must be included in the model to account for the experimentally measured values of quadrupole energies $E_{QZ}$, $E_{QR}$ and $E_{QI}$.
We end up with the in-plane pressure parameters $p_1=-0.5$~MP, $p_2=2.5$~MP, $\varsigma=-1.18$, and electric field $F=5$~kV/cm. 

Although the presence of  electric fields close to the surface is not  impossible due to the pinning of the Fermi level on the surface states in the middle of the GaAs bandgap \cite{PhysRevB.43.12138}, such  electric field could, in principle, ionize the donors, preventing dynamic polarization of nuclear spins by localized electrons  \cite{PhysRevB.76.245301,King2012}. 
Moreover, we have conducted EBIC characterization of the sample.
These results, presented in Appendix suggest that there is no measurable surface-induced electric field.
The only region where an electric field is detected by EBIC technique is located at $77$~$\muup$m from the surface, close to the interface between $n$-GaAs layer and $p$-GaAs substrate. 
%
%

In order to describe the warm-up spectra by the deformation induced quadrupole effects only we need to admit that the gradient-elastic tensor element  $S_{44}$ for $^{75}$As  is  different from  the value estimated in Ref.~\onlinecite{Griffiths2019}. The resulting values of $S_{44}$ tensor elements are given in Table 1.
One can see that the difference with Ref.~\onlinecite{Griffiths2019}  is significant, exceeding a factor of two. 
The resulting stress estimation yields  $p_1=6.5$~MP, $p_2=-8.5$~MP and $\varsigma=-0.73$. This corresponds to both tensile/compression of the same order as measured previously in bulk GaAs samples at low temperatures,  and $\approx 5$ times  more significant shear strain \cite{Yamada1998l}. 
We speculate that mechanical deformations can be associated with different thermal expansion coefficients for GaAs sample and sapphire holder at $T=$ 20 K. Also, glue between the sample and the sample holder can induce some uncontrollable squeezing. 
%
%

\section{Conclusions}

In conclusion, we have shown that nuclear spin temperature can be used as a probe of the NSS state in the presence of an oscillating magnetic field at various frequencies.  This method is termed warm-up spectroscopy, because it is particularly fruitful at low magnetic fields, where magnetization of the NSS is low, and traditional optically detected NMR does not have enough sensitivity. The warm-up spectroscopy addresses precooled NSS to increase its sensitivity. The degree of heating of the NSS by the OMF in the dark is subsequently determined via the Overhauser field.  

We measured NMR spectra detected by such a thermometer for different mutual orientations of static external magnetic field and OMF, and with respect to \textit{n}-GaAs epilayer crystal axes.  The analysis of the data within the model accounting for various sources of the quadrupole interaction and the NSS parameters from the literature suggests that  both mechanical strain and built-in electric field contribute to the quadrupole splitting of different isotopes. Nevertheless, since we could not evidence the presence of such electric field, the  off-diagonal element of the  $^{75}$As gradient-electric tensor may need to be revisited. We obtained $S_{44}=(17\pm 3) \times10^{21}$~V/m$^2$, that is more than two times less than in Ref.~\onlinecite{Griffiths2019}.  Thus, further experiments with in-situ control of the electric field and deformation will be a key to better understanding of their relative contributions and determination of the gradient-elastic tensors for $^{75}$As. %
Overall, our results validate warm-up spectroscopy as a new effective tool for studies of the quadrupole-split NSS.

\section{Acknowledgements}

The authors are grateful to M.~M.~Sobolev for inspiring discussions and  acknowledge Saint-Petersburg State University for a research grant 73031758, financial support from the Russian Foundation for Basic Research (RFBR Project No. 19-52-12043) and Deutsche Forschungsgemeinschaft (DFG Project A6) in the framework of International Collaborative Research Center TRR 160. 
{SEM and} EBIC characterization was performed using equipment owned by the Federal Joint Research Center "Material science and characterization in advanced technology".
V.M.L. acknowledges Russian Foundation for Basic Research for research Grant 19-32-90084 and the support of the French Embassy in Moscow (Ostrogradski fellowship for young researchers 2020).

\section{Appendix : Estimation of the built-in  electric field by EBIC microscopy}
\label{sec:appendix}
\begin{figure}
\includegraphics[width=3.2in]{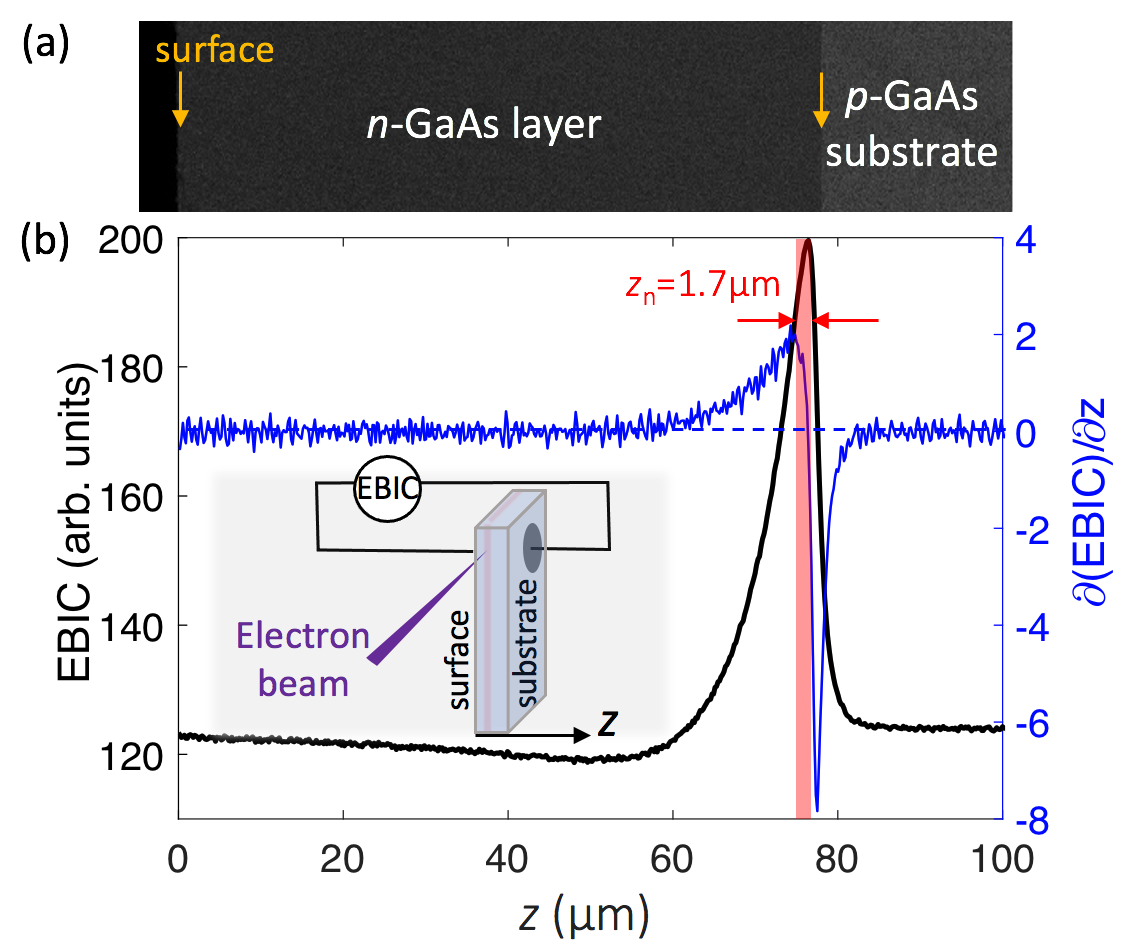}
\caption{(a) {SEM image of sample in cross-sectional geometry.  Variations of the grey shades (also marked by yellow arrows) point the positions of the sample surface, as well as the interface between the epitaxial layer and $p$-GaAs substrate.}
(b) Electron beam induced current (black line) and derivative of this current (blue line) as a function of the beam position on the cleaved sample edge. The electron beam energy is  $15$~keV.
Pink area shows the SCR on the $n$-side of the $p-n$ junction. It expands from the position corresponding to the  EBIC maximum (zero derivative) to the position of the EBIC inflection point (maximum of the derivative). SCR  thickness is $z_n=1.7$~$\muup$m. 
Inset shows the cross-sectional geometry of the scan, where electron beam is perpendicular to the cleaved edge of the sample.  }
\label{fig:EBIC}
\end{figure}
EBIC  is a powerful technique based on the collection of free charges generated within a semiconductor by an electron beam. 
Cross-sectional EBIC  is routinely applied to identify the location of $p$–$n$ or Schottky junctions, as well as the corresponding built-in electron fields \cite{FRIGERI20012557,Oelgart}. In the setup used in this work it is  integrated in a scanning electron microscope (SEM).
%

The sketch of the cross-sectional EBIC experiment, SEM image of the sample cross-section,  and the measured current  are shown in Fig.~\ref{fig:EBIC}. 
By analysing the contrast of the SEM image one can  identify the surface of the sample,  $n$-GaAs layer, as well as $p$-GaAs substrate at $77$~$\muup$m from the surface. 
The  peak of the current  corresponds to increased electric field in the space charge region (SCR) on both sides of the $p$-$n$ junction, situated close to the interface with the substrate. 
By contrast,  there is no measurable variation of the signal in the vicinity of the sample surface, in the region probed by warm-up spectroscopy.  
%
%

In order to estimate the maximum electric field of the $p-n$ junction $F_{p-n}$, we first determine the  SCR width on the $n$-doped side of the junction. It is defined as the distance between the positions of the current maximum (zero derivative) and the inflection point (maximum derivative), which yields {$z_n=1.7$~$\muup$m}, pink area in Fig.~\ref{fig:EBIC}.
Then, the maximum electric field is given by \cite{Luth2015}
\begin{equation} 
\label{eq:EBIC} 
F_{p-n}=-\frac{e n_d z_n}{\epsilon \epsilon_0},
\end{equation} 
where $\epsilon=12.9$ is GaAs dielectric permittivity,  $\epsilon_0$ is vacuum permittivity.  Assuming nominal donor density $n_d=10^{15}$~cm$^{-3}$ (consistent with measured electron and nuclear spin relaxation times)  we obtain $F_{p-n}\approx 24$~kV/cm.
In contrast to the interface region, the EBIC signal has no peculiarities near the surface, suggesting that any field $F$ near the surface is less than $F_{p-n}$, if existing at all. One can reasonably assume that it does not exceed $1$~kV/cm. Thus  we conclude that the electric field plays a minor role, while the deformation is the dominant source of the quadrupole splittings in the studied sample.

\bibliography{refs}

\begin{thebibliography}{46}%
\makeatletter
\providecommand \@ifxundefined [1]{%
 \@ifx{#1\undefined}
}%
\providecommand \@ifnum [1]{%
 \ifnum #1\expandafter \@firstoftwo
 \else \expandafter \@secondoftwo
 \fi
}%
\providecommand \@ifx [1]{%
 \ifx #1\expandafter \@firstoftwo
 \else \expandafter \@secondoftwo
 \fi
}%
\providecommand \natexlab [1]{#1}%
\providecommand \enquote  [1]{``#1''}%
\providecommand \bibnamefont  [1]{#1}%
\providecommand \bibfnamefont [1]{#1}%
\providecommand \citenamefont [1]{#1}%
\providecommand \href@noop [0]{\@secondoftwo}%
\providecommand \href [0]{\begingroup \@sanitize@url \@href}%
\providecommand \@href[1]{\@@startlink{#1}\@@href}%
\providecommand \@@href[1]{\endgroup#1\@@endlink}%
\providecommand \@sanitize@url [0]{\catcode `\\12\catcode `\$12\catcode
  `\&12\catcode `\#12\catcode `\^12\catcode `\_12\catcode `\%12\relax}%
\providecommand \@@startlink[1]{}%
\providecommand \@@endlink[0]{}%
\providecommand \url  [0]{\begingroup\@sanitize@url \@url }%
\providecommand \@url [1]{\endgroup\@href {#1}{\urlprefix }}%
\providecommand \urlprefix  [0]{URL }%
\providecommand \Eprint [0]{\href }%
\providecommand \doibase [0]{https://doi.org/}%
\providecommand \selectlanguage [0]{\@gobble}%
\providecommand \bibinfo  [0]{\@secondoftwo}%
\providecommand \bibfield  [0]{\@secondoftwo}%
\providecommand \translation [1]{[#1]}%
\providecommand \BibitemOpen [0]{}%
\providecommand \bibitemStop [0]{}%
\providecommand \bibitemNoStop [0]{.\EOS\space}%
\providecommand \EOS [0]{\spacefactor3000\relax}%
\providecommand \BibitemShut  [1]{\csname bibitem#1\endcsname}%
\let\auto@bib@innerbib\@empty
\bibitem [{\citenamefont {Meier}\ and\ \citenamefont
  {Zakharchenya}(1984)}]{OO1}%
  \BibitemOpen
  \bibinfo {editor} {\bibfnamefont {F.}~\bibnamefont {Meier}}\ and\ \bibinfo
  {editor} {\bibfnamefont {B.~P.}\ \bibnamefont {Zakharchenya}},\ eds.,\
  \href@noop {} {\emph {\bibinfo {title} {Optical Orientation}}}\ (\bibinfo
  {publisher} {North Holland},\ \bibinfo {address} {Amsterdam},\ \bibinfo
  {year} {1984})\BibitemShut {NoStop}%
\bibitem [{\citenamefont {Kotur}\ \emph {et~al.}(2016)\citenamefont {Kotur},
  \citenamefont {Dzhioev}, \citenamefont {Vladimirova}, \citenamefont
  {Jouault}, \citenamefont {Korenev},\ and\ \citenamefont
  {Kavokin}}]{PhysRevB.94.081201}%
  \BibitemOpen
  \bibfield  {author} {\bibinfo {author} {\bibfnamefont {M.}~\bibnamefont
  {Kotur}}, \bibinfo {author} {\bibfnamefont {R.~I.}\ \bibnamefont {Dzhioev}},
  \bibinfo {author} {\bibfnamefont {M.}~\bibnamefont {Vladimirova}}, \bibinfo
  {author} {\bibfnamefont {B.}~\bibnamefont {Jouault}}, \bibinfo {author}
  {\bibfnamefont {V.~L.}\ \bibnamefont {Korenev}},\ and\ \bibinfo {author}
  {\bibfnamefont {K.~V.}\ \bibnamefont {Kavokin}},\ }\bibfield  {title}
  {\bibinfo {title} {Nuclear spin warm up in bulk $n$-{GaAs}},\ }\href
  {https://doi.org/10.1103/PhysRevB.94.081201} {\bibfield  {journal} {\bibinfo
  {journal} {Phys. Rev. B}\ }\textbf {\bibinfo {volume} {94}},\ \bibinfo
  {pages} {081201} (\bibinfo {year} {2016})}\BibitemShut {NoStop}%
\bibitem [{\citenamefont {Vladimirova}\ \emph {et~al.}(2017)\citenamefont
  {Vladimirova}, \citenamefont {Cronenberger}, \citenamefont {Scalbert},
  \citenamefont {Kotur}, \citenamefont {Dzhioev}, \citenamefont {Ryzhov},
  \citenamefont {Kozlov}, \citenamefont {Zapasskii}, \citenamefont
  {Lema\^{\i}tre},\ and\ \citenamefont {Kavokin}}]{PhysRevB.95.125312}%
  \BibitemOpen
  \bibfield  {author} {\bibinfo {author} {\bibfnamefont {M.}~\bibnamefont
  {Vladimirova}}, \bibinfo {author} {\bibfnamefont {S.}~\bibnamefont
  {Cronenberger}}, \bibinfo {author} {\bibfnamefont {D.}~\bibnamefont
  {Scalbert}}, \bibinfo {author} {\bibfnamefont {M.}~\bibnamefont {Kotur}},
  \bibinfo {author} {\bibfnamefont {R.~I.}\ \bibnamefont {Dzhioev}}, \bibinfo
  {author} {\bibfnamefont {I.~I.}\ \bibnamefont {Ryzhov}}, \bibinfo {author}
  {\bibfnamefont {G.~G.}\ \bibnamefont {Kozlov}}, \bibinfo {author}
  {\bibfnamefont {V.~S.}\ \bibnamefont {Zapasskii}}, \bibinfo {author}
  {\bibfnamefont {A.}~\bibnamefont {Lema\^{\i}tre}},\ and\ \bibinfo {author}
  {\bibfnamefont {K.~V.}\ \bibnamefont {Kavokin}},\ }\bibfield  {title}
  {\bibinfo {title} {Nuclear spin relaxation in $n$-{GaAs}: From insulating to
  metallic regime},\ }\href {https://doi.org/10.1103/PhysRevB.95.125312}
  {\bibfield  {journal} {\bibinfo  {journal} {Phys. Rev. B}\ }\textbf {\bibinfo
  {volume} {95}},\ \bibinfo {pages} {125312} (\bibinfo {year}
  {2017})}\BibitemShut {NoStop}%
\bibitem [{\citenamefont {Abragam}\ and\ \citenamefont
  {Proctor}(1958)}]{AbragamProctor}%
  \BibitemOpen
  \bibfield  {author} {\bibinfo {author} {\bibfnamefont {A.}~\bibnamefont
  {Abragam}}\ and\ \bibinfo {author} {\bibfnamefont {W.~G.}\ \bibnamefont
  {Proctor}},\ }\bibfield  {title} {\bibinfo {title} {{Spin temperature}},\
  }\href {https://doi.org/10.1103/PhysRev.109.1441} {\bibfield  {journal}
  {\bibinfo  {journal} {Phys. Rev.}\ }\textbf {\bibinfo {volume} {109}},\
  \bibinfo {pages} {1441} (\bibinfo {year} {1958})}\BibitemShut {NoStop}%
\bibitem [{\citenamefont {Merkulov}(1982)}]{Merkulov1982}%
  \BibitemOpen
  \bibfield  {author} {\bibinfo {author} {\bibfnamefont {I.~A.}\ \bibnamefont
  {Merkulov}},\ }\bibfield  {title} {\bibinfo {title} {{Phase transition in the
  nuclear spin system in a semiconductor with optically oriented electrons}},\
  }\href@noop {} {\bibfield  {journal} {\bibinfo  {journal} {Soviet Physics
  JETP}\ }\textbf {\bibinfo {volume} {55}},\ \bibinfo {pages} {188} (\bibinfo
  {year} {1982})}\BibitemShut {NoStop}%
\bibitem [{\citenamefont {Merkulov}\ \emph {et~al.}(1987)\citenamefont
  {Merkulov}, \citenamefont {Papava}, \citenamefont {Ponomarenko},\ and\
  \citenamefont {Vasiliev}}]{Merkulov1987}%
  \BibitemOpen
  \bibfield  {author} {\bibinfo {author} {\bibfnamefont {I.~A.}\ \bibnamefont
  {Merkulov}}, \bibinfo {author} {\bibfnamefont {Y.~I.}\ \bibnamefont
  {Papava}}, \bibinfo {author} {\bibfnamefont {V.~V.}\ \bibnamefont
  {Ponomarenko}},\ and\ \bibinfo {author} {\bibfnamefont {S.~I.}\ \bibnamefont
  {Vasiliev}},\ }\bibfield  {title} {\bibinfo {title} {{Monte Carlo simulation
  and theory in Gaussian approximation of a phase transition in the nuclear
  spin system of a solid}},\ }\href@noop {} {\bibfield  {journal} {\bibinfo
  {journal} {Canadian Journal of Physics}\ }\textbf {\bibinfo {volume} {66}},\
  \bibinfo {pages} {135} (\bibinfo {year} {1987})}\BibitemShut {NoStop}%
\bibitem [{\citenamefont {Merkulov}(1998)}]{Merkulov1998}%
  \BibitemOpen
  \bibfield  {author} {\bibinfo {author} {\bibfnamefont {I.}~\bibnamefont
  {Merkulov}},\ }\bibfield  {title} {\bibinfo {title} {Formation of a nuclear
  spin polaron under optical orientation in {GaAs}-type semiconductors},\
  }\href@noop {} {\bibfield  {journal} {\bibinfo  {journal} {Physics of the
  Solid State}\ }\textbf {\bibinfo {volume} {40}},\ \bibinfo {pages} {930}
  (\bibinfo {year} {1998})}\BibitemShut {NoStop}%
\bibitem [{\citenamefont {Vladimirova}\ \emph {et~al.}(2021)\citenamefont
  {Vladimirova}, \citenamefont {Scalbert}, \citenamefont {Kuznetsova},\ and\
  \citenamefont {Kavokin}}]{vladimirova2021}%
  \BibitemOpen
  \bibfield  {author} {\bibinfo {author} {\bibfnamefont {M.}~\bibnamefont
  {Vladimirova}}, \bibinfo {author} {\bibfnamefont {D.}~\bibnamefont
  {Scalbert}}, \bibinfo {author} {\bibfnamefont {M.~S.}\ \bibnamefont
  {Kuznetsova}},\ and\ \bibinfo {author} {\bibfnamefont {K.~V.}\ \bibnamefont
  {Kavokin}},\ }\bibfield  {title} {\bibinfo {title} {{Electron-induced nuclear
  magnetic ordering in $n$-type semiconductors}},\ }\href@noop {} {\bibfield
  {journal} {\bibinfo  {journal} {Phys. Rev. B}\ }\textbf {\bibinfo {volume}
  {103}},\ \bibinfo {pages} {205207} (\bibinfo {year} {2021})}\BibitemShut
  {NoStop}%
\bibitem [{\citenamefont {Lampel}(1968)}]{PhysRevLett.20.491}%
  \BibitemOpen
  \bibfield  {author} {\bibinfo {author} {\bibfnamefont {G.}~\bibnamefont
  {Lampel}},\ }\bibfield  {title} {\bibinfo {title} {Nuclear dynamic
  polarization by optical electronic saturation and optical pumping in
  semiconductors},\ }\href {https://doi.org/10.1103/PhysRevLett.20.491}
  {\bibfield  {journal} {\bibinfo  {journal} {Phys. Rev. Lett.}\ }\textbf
  {\bibinfo {volume} {20}},\ \bibinfo {pages} {491} (\bibinfo {year}
  {1968})}\BibitemShut {NoStop}%
\bibitem [{\citenamefont {Kalevich}\ \emph {et~al.}(1982)\citenamefont
  {Kalevich}, \citenamefont {Kulkov},\ and\ \citenamefont
  {Fleisher}}]{JETP1982}%
  \BibitemOpen
  \bibfield  {author} {\bibinfo {author} {\bibfnamefont {V.~K.}\ \bibnamefont
  {Kalevich}}, \bibinfo {author} {\bibfnamefont {V.~D.}\ \bibnamefont
  {Kulkov}},\ and\ \bibinfo {author} {\bibfnamefont {V.~G.}\ \bibnamefont
  {Fleisher}},\ }\bibfield  {title} {\bibinfo {title} {Onset of a nuclear
  polarization front due to optical spin orientation in a semiconductor},\
  }\href@noop {} {\bibfield  {journal} {\bibinfo  {journal} {JETP Lett.}\
  }\textbf {\bibinfo {volume} {35}},\ \bibinfo {pages} {20} (\bibinfo {year}
  {1982})}\BibitemShut {NoStop}%
\bibitem [{\citenamefont {Kalevich}\ \emph {et~al.}(2017)\citenamefont
  {Kalevich}, \citenamefont {Kavokin}, \citenamefont {Merkulov},\ and\
  \citenamefont {Vladimirova}}]{CH11Spin2017}%
  \BibitemOpen
  \bibfield  {author} {\bibinfo {author} {\bibfnamefont {V.~K.}\ \bibnamefont
  {Kalevich}}, \bibinfo {author} {\bibfnamefont {K.~V.}\ \bibnamefont
  {Kavokin}}, \bibinfo {author} {\bibfnamefont {I.}~\bibnamefont {Merkulov}},\
  and\ \bibinfo {author} {\bibfnamefont {M.}~\bibnamefont {Vladimirova}},\
  }\bibfield  {title} {\bibinfo {title} {{Dynamic Nuclear Polarization and
  Nuclear Fields}},\ }in\ \href {https://doi.org/10.1007/978-3-319-65436-2_12}
  {\emph {\bibinfo {booktitle} {Spin Physics in Semiconductors}}},\ \bibinfo
  {editor} {edited by\ \bibinfo {editor} {\bibfnamefont {M.~I.}\ \bibnamefont
  {Dyakonov}}}\ (\bibinfo  {publisher} {Springer International Publishing},\
  \bibinfo {address} {Cham},\ \bibinfo {year} {2017})\ pp.\ \bibinfo {pages}
  {387--430}\BibitemShut {NoStop}%
\bibitem [{\citenamefont {Vladimirova}\ \emph {et~al.}(2018)\citenamefont
  {Vladimirova}, \citenamefont {Cronenberger}, \citenamefont {Scalbert},
  \citenamefont {Ryzhov}, \citenamefont {Zapasskii}, \citenamefont {Kozlov},
  \citenamefont {Lema\^{\i}tre},\ and\ \citenamefont
  {Kavokin}}]{PhysRevB.97.041301}%
  \BibitemOpen
  \bibfield  {author} {\bibinfo {author} {\bibfnamefont {M.}~\bibnamefont
  {Vladimirova}}, \bibinfo {author} {\bibfnamefont {S.}~\bibnamefont
  {Cronenberger}}, \bibinfo {author} {\bibfnamefont {D.}~\bibnamefont
  {Scalbert}}, \bibinfo {author} {\bibfnamefont {I.~I.}\ \bibnamefont
  {Ryzhov}}, \bibinfo {author} {\bibfnamefont {V.~S.}\ \bibnamefont
  {Zapasskii}}, \bibinfo {author} {\bibfnamefont {G.~G.}\ \bibnamefont
  {Kozlov}}, \bibinfo {author} {\bibfnamefont {A.}~\bibnamefont
  {Lema\^{\i}tre}},\ and\ \bibinfo {author} {\bibfnamefont {K.~V.}\
  \bibnamefont {Kavokin}},\ }\bibfield  {title} {\bibinfo {title} {Spin
  temperature concept verified by optical magnetometry of nuclear spins},\
  }\href {https://doi.org/10.1103/PhysRevB.97.041301} {\bibfield  {journal}
  {\bibinfo  {journal} {Phys. Rev. B}\ }\textbf {\bibinfo {volume} {97}},\
  \bibinfo {pages} {041301} (\bibinfo {year} {2018})}\BibitemShut {NoStop}%
\bibitem [{\citenamefont {Kotur}\ \emph {et~al.}(2021)\citenamefont {Kotur},
  \citenamefont {Tolmachev}, \citenamefont {Litvyak}, \citenamefont {Kavokin},
  \citenamefont {Suter}, \citenamefont {Yakovlev},\ and\ \citenamefont
  {Bayer}}]{kotur2021deep}%
  \BibitemOpen
  \bibfield  {author} {\bibinfo {author} {\bibfnamefont {M.}~\bibnamefont
  {Kotur}}, \bibinfo {author} {\bibfnamefont {D.~O.}\ \bibnamefont
  {Tolmachev}}, \bibinfo {author} {\bibfnamefont {V.~M.}\ \bibnamefont
  {Litvyak}}, \bibinfo {author} {\bibfnamefont {K.~V.}\ \bibnamefont
  {Kavokin}}, \bibinfo {author} {\bibfnamefont {D.}~\bibnamefont {Suter}},
  \bibinfo {author} {\bibfnamefont {D.~R.}\ \bibnamefont {Yakovlev}},\ and\
  \bibinfo {author} {\bibfnamefont {M.}~\bibnamefont {Bayer}},\ }\bibfield
  {title} {\bibinfo {title} {Ultra-deep optical cooling of coupled nuclear
  spin-spin and quadrupole reservoirs in a {GaAs/(Al,Ga)As}f quantum well},\
  }\href {https://doi.org/10.1038/s42005-021-00681-6} {\bibfield  {journal}
  {\bibinfo  {journal} {Communications Physics}\ }\textbf {\bibinfo {volume}
  {4}},\ \bibinfo {pages} {193} (\bibinfo {year} {2021})}\BibitemShut {NoStop}%
\bibitem [{\citenamefont {Goldman}(1970)}]{Goldman}%
  \BibitemOpen
  \bibfield  {author} {\bibinfo {author} {\bibfnamefont {M.}~\bibnamefont
  {Goldman}},\ }\href@noop {} {\emph {\bibinfo {title} {Spin Temperature and
  Nuclear Magnetic Resonance in Solids}}}\ (\bibinfo  {publisher} {Oxford
  University Press},\ \bibinfo {year} {1970})\BibitemShut {NoStop}%
\bibitem [{\citenamefont {Poggio}\ and\ \citenamefont
  {Awschalom}(2005)}]{doi:10.1063/1.1923191}%
  \BibitemOpen
  \bibfield  {author} {\bibinfo {author} {\bibfnamefont {M.}~\bibnamefont
  {Poggio}}\ and\ \bibinfo {author} {\bibfnamefont {D.~D.}\ \bibnamefont
  {Awschalom}},\ }\bibfield  {title} {\bibinfo {title} {High-field optically
  detected nuclear magnetic resonance in {GaAs}},\ }\href
  {https://doi.org/10.1063/1.1923191} {\bibfield  {journal} {\bibinfo
  {journal} {Applied. Phys. Lett.}\ }\textbf {\bibinfo {volume} {86}},\
  \bibinfo {pages} {182103} (\bibinfo {year} {2005})}\BibitemShut {NoStop}%
\bibitem [{\citenamefont {Eickhoff}\ \emph {et~al.}(2003)\citenamefont
  {Eickhoff}, \citenamefont {Lenzmann}, \citenamefont {Suter}, \citenamefont
  {Hayes},\ and\ \citenamefont {Wieck}}]{PhysRevB.67.085308}%
  \BibitemOpen
  \bibfield  {author} {\bibinfo {author} {\bibfnamefont {M.}~\bibnamefont
  {Eickhoff}}, \bibinfo {author} {\bibfnamefont {B.}~\bibnamefont {Lenzmann}},
  \bibinfo {author} {\bibfnamefont {D.}~\bibnamefont {Suter}}, \bibinfo
  {author} {\bibfnamefont {S.~E.}\ \bibnamefont {Hayes}},\ and\ \bibinfo
  {author} {\bibfnamefont {A.~D.}\ \bibnamefont {Wieck}},\ }\bibfield  {title}
  {\bibinfo {title} {Mapping of strain and electric fields in
  {${\mathrm{G}\mathrm{a}\mathrm{A}\mathrm{s}/\mathrm{A}\mathrm{l}}_{x}{\mathrm{Ga}}_{1\ensuremath{-}x}\mathrm{As}$}
  quantum-well samples by laser-assisted {NMR}},\ }\href
  {https://doi.org/10.1103/PhysRevB.67.085308} {\bibfield  {journal} {\bibinfo
  {journal} {Phys. Rev. B}\ }\textbf {\bibinfo {volume} {67}},\ \bibinfo
  {pages} {085308} (\bibinfo {year} {2003})}\BibitemShut {NoStop}%
\bibitem [{\citenamefont {{Maletinsky}}\ \emph {et~al.}(2009)\citenamefont
  {{Maletinsky}}, \citenamefont {{Kroner}},\ and\ \citenamefont
  {{Imamo\u{g}lu}}}]{Nat09}%
  \BibitemOpen
  \bibfield  {author} {\bibinfo {author} {\bibfnamefont {P.}~\bibnamefont
  {{Maletinsky}}}, \bibinfo {author} {\bibfnamefont {M.}~\bibnamefont
  {{Kroner}}},\ and\ \bibinfo {author} {\bibfnamefont {A.}~\bibnamefont
  {{Imamo\u{g}lu}}},\ }\bibfield  {title} {\bibinfo {title} {Breakdown of the
  nuclear-spin-temperature approach in quantum-dot demagnetization
  experiments},\ }\href@noop {} {\bibfield  {journal} {\bibinfo  {journal}
  {Nat. Phys.}\ }\textbf {\bibinfo {volume} {5}},\ \bibinfo {pages} {407}
  (\bibinfo {year} {2009})}\BibitemShut {NoStop}%
\bibitem [{\citenamefont {Chekhovich}\ \emph {et~al.}(2012)\citenamefont
  {Chekhovich}, \citenamefont {Kavokin}, \citenamefont {Puebla}, \citenamefont
  {Krysa}, \citenamefont {Hopkinson}, \citenamefont {Andreev}, \citenamefont
  {Sanchez}, \citenamefont {Beanland}, \citenamefont {Skolnick},\ and\
  \citenamefont {Tartakovskii}}]{Chekhovich2012}%
  \BibitemOpen
  \bibfield  {author} {\bibinfo {author} {\bibfnamefont {E.~A.}\ \bibnamefont
  {Chekhovich}}, \bibinfo {author} {\bibfnamefont {K.~V.}\ \bibnamefont
  {Kavokin}}, \bibinfo {author} {\bibfnamefont {J.}~\bibnamefont {Puebla}},
  \bibinfo {author} {\bibfnamefont {A.~B.}\ \bibnamefont {Krysa}}, \bibinfo
  {author} {\bibfnamefont {M.}~\bibnamefont {Hopkinson}}, \bibinfo {author}
  {\bibfnamefont {A.~D.}\ \bibnamefont {Andreev}}, \bibinfo {author}
  {\bibfnamefont {A.~M.}\ \bibnamefont {Sanchez}}, \bibinfo {author}
  {\bibfnamefont {R.}~\bibnamefont {Beanland}}, \bibinfo {author}
  {\bibfnamefont {M.~S.}\ \bibnamefont {Skolnick}},\ and\ \bibinfo {author}
  {\bibfnamefont {A.~I.}\ \bibnamefont {Tartakovskii}},\ }\bibfield  {title}
  {\bibinfo {title} {Structural analysis of strained quantum dots using nuclear
  magnetic resonance},\ }\href {https://doi.org/10.1038/nnano.2012.142}
  {\bibfield  {journal} {\bibinfo  {journal} {Nat. Nanotech.}\ }\textbf
  {\bibinfo {volume} {7}},\ \bibinfo {pages} {646} (\bibinfo {year}
  {2012})}\BibitemShut {NoStop}%
\bibitem [{\citenamefont {Chekhovich}\ \emph {et~al.}(2017)\citenamefont
  {Chekhovich}, \citenamefont {Ulhaq}, \citenamefont {Zallo}, \citenamefont
  {Ding}, \citenamefont {Schmidt},\ and\ \citenamefont
  {Skolnick}}]{Chekhovich2017}%
  \BibitemOpen
  \bibfield  {author} {\bibinfo {author} {\bibfnamefont {E.~A.}\ \bibnamefont
  {Chekhovich}}, \bibinfo {author} {\bibfnamefont {A.}~\bibnamefont {Ulhaq}},
  \bibinfo {author} {\bibfnamefont {E.}~\bibnamefont {Zallo}}, \bibinfo
  {author} {\bibfnamefont {F.}~\bibnamefont {Ding}}, \bibinfo {author}
  {\bibfnamefont {O.~G.}\ \bibnamefont {Schmidt}},\ and\ \bibinfo {author}
  {\bibfnamefont {M.~S.}\ \bibnamefont {Skolnick}},\ }\bibfield  {title}
  {\bibinfo {title} {Measurement of the spin temperature of optically cooled
  nuclei and {GaAs} hyperfine constants in {GaAs/AlGaAs} quantum dots},\ }\href
  {https://doi.org/10.1038/nmat4959} {\bibfield  {journal} {\bibinfo  {journal}
  {Nat. Mat.}\ }\textbf {\bibinfo {volume} {16}},\ \bibinfo {pages} {982}
  (\bibinfo {year} {2017})}\BibitemShut {NoStop}%
\bibitem [{\citenamefont {Barrett}\ \emph {et~al.}(1994)\citenamefont
  {Barrett}, \citenamefont {Tycko}, \citenamefont {Pfeiffer},\ and\
  \citenamefont {West}}]{PhysRevLett.72.1368}%
  \BibitemOpen
  \bibfield  {author} {\bibinfo {author} {\bibfnamefont {S.~E.}\ \bibnamefont
  {Barrett}}, \bibinfo {author} {\bibfnamefont {R.}~\bibnamefont {Tycko}},
  \bibinfo {author} {\bibfnamefont {L.~N.}\ \bibnamefont {Pfeiffer}},\ and\
  \bibinfo {author} {\bibfnamefont {K.~W.}\ \bibnamefont {West}},\ }\bibfield
  {title} {\bibinfo {title} {Directly detected nuclear magnetic resonance of
  optically pumped {GaAs} quantum wells},\ }\href
  {https://doi.org/10.1103/PhysRevLett.72.1368} {\bibfield  {journal} {\bibinfo
   {journal} {Phys. Rev. Lett.}\ }\textbf {\bibinfo {volume} {72}},\ \bibinfo
  {pages} {1368} (\bibinfo {year} {1994})}\BibitemShut {NoStop}%
\bibitem [{\citenamefont {Litvyak}\ \emph {et~al.}(2018)\citenamefont
  {Litvyak}, \citenamefont {Cherbunin}, \citenamefont {Kavokin},\ and\
  \citenamefont {Kalevich}}]{Lit18}%
  \BibitemOpen
  \bibfield  {author} {\bibinfo {author} {\bibfnamefont {V.~M.}\ \bibnamefont
  {Litvyak}}, \bibinfo {author} {\bibfnamefont {R.}~\bibnamefont {Cherbunin}},
  \bibinfo {author} {\bibfnamefont {K.}~\bibnamefont {Kavokin}},\ and\ \bibinfo
  {author} {\bibfnamefont {V.}~\bibnamefont {Kalevich}},\ }\bibfield  {title}
  {\bibinfo {title} {Determination of the local field in the nuclear spin
  system of \textit{n}-type {GaAs}},\ }\href@noop {} {\bibfield  {journal}
  {\bibinfo  {journal} {IOP Conf. Series: Journal of Physics: Conference
  Series}\ }\textbf {\bibinfo {volume} {951}},\ \bibinfo {pages} {012006}
  (\bibinfo {year} {2018})}\BibitemShut {NoStop}%
\bibitem [{\citenamefont {Paget}\ \emph {et~al.}(1977)\citenamefont {Paget},
  \citenamefont {Lampel}, \citenamefont {Sapoval},\ and\ \citenamefont
  {Safarov}}]{PhysRevB.15.5780}%
  \BibitemOpen
  \bibfield  {author} {\bibinfo {author} {\bibfnamefont {D.}~\bibnamefont
  {Paget}}, \bibinfo {author} {\bibfnamefont {G.}~\bibnamefont {Lampel}},
  \bibinfo {author} {\bibfnamefont {B.}~\bibnamefont {Sapoval}},\ and\ \bibinfo
  {author} {\bibfnamefont {V.~I.}\ \bibnamefont {Safarov}},\ }\bibfield
  {title} {\bibinfo {title} {Low field electron-nuclear spin coupling in
  gallium arsenide under optical pumping conditions},\ }\href
  {https://doi.org/10.1103/PhysRevB.15.5780} {\bibfield  {journal} {\bibinfo
  {journal} {Phys. Rev. B}\ }\textbf {\bibinfo {volume} {15}},\ \bibinfo
  {pages} {5780} (\bibinfo {year} {1977})}\BibitemShut {NoStop}%
\bibitem [{\citenamefont {Kempf}\ \emph {et~al.}(2008)\citenamefont {Kempf},
  \citenamefont {Marohn}, \citenamefont {Carson}, \citenamefont {Shykind},
  \citenamefont {Hwang}, \citenamefont {Miller},\ and\ \citenamefont
  {Weitekamp}}]{RSI08}%
  \BibitemOpen
  \bibfield  {author} {\bibinfo {author} {\bibfnamefont {J.~G.}\ \bibnamefont
  {Kempf}}, \bibinfo {author} {\bibfnamefont {J.~A.}\ \bibnamefont {Marohn}},
  \bibinfo {author} {\bibfnamefont {P.~J.}\ \bibnamefont {Carson}}, \bibinfo
  {author} {\bibfnamefont {D.~A.}\ \bibnamefont {Shykind}}, \bibinfo {author}
  {\bibfnamefont {J.~Y.}\ \bibnamefont {Hwang}}, \bibinfo {author}
  {\bibfnamefont {M.~A.}\ \bibnamefont {Miller}},\ and\ \bibinfo {author}
  {\bibfnamefont {D.~P.}\ \bibnamefont {Weitekamp}},\ }\bibfield  {title}
  {\bibinfo {title} {{An optical NMR spectrometer for Larmor-beat detection and
  high-resolution POWER NMR}},\ }\href {https://doi.org/10.1063/1.2936257}
  {\bibfield  {journal} {\bibinfo  {journal} {Review of Scientific
  Instruments}\ }\textbf {\bibinfo {volume} {79}},\ \bibinfo {pages} {063904}
  (\bibinfo {year} {2008})}\BibitemShut {NoStop}%
\bibitem [{\citenamefont {Kotur}\ \emph {et~al.}(2014)\citenamefont {Kotur},
  \citenamefont {Dzhioev}, \citenamefont {Kavokin}, \citenamefont {Korenev},
  \citenamefont {Namozov}, \citenamefont {Pak},\ and\ \citenamefont
  {Kusrayev}}]{JETP99}%
  \BibitemOpen
  \bibfield  {author} {\bibinfo {author} {\bibfnamefont {M.}~\bibnamefont
  {Kotur}}, \bibinfo {author} {\bibfnamefont {R.~I.}\ \bibnamefont {Dzhioev}},
  \bibinfo {author} {\bibfnamefont {K.~V.}\ \bibnamefont {Kavokin}}, \bibinfo
  {author} {\bibfnamefont {V.~L.}\ \bibnamefont {Korenev}}, \bibinfo {author}
  {\bibfnamefont {B.~R.}\ \bibnamefont {Namozov}}, \bibinfo {author}
  {\bibfnamefont {P.~E.}\ \bibnamefont {Pak}},\ and\ \bibinfo {author}
  {\bibfnamefont {Y.~G.}\ \bibnamefont {Kusrayev}},\ }\bibfield  {title}
  {\bibinfo {title} {Nuclear spin relaxation mediated by fermi-edge electrons
  in \textit{n}-type {GaAs}},\ }\href@noop {} {\bibfield  {journal} {\bibinfo
  {journal} {JETP Lett.}\ }\textbf {\bibinfo {volume} {99}},\ \bibinfo {pages}
  {37} (\bibinfo {year} {2014})}\BibitemShut {NoStop}%
\bibitem [{\citenamefont {Kalevich}\ and\ \citenamefont
  {Fleisher}(1983)}]{Bill83}%
  \BibitemOpen
  \bibfield  {author} {\bibinfo {author} {\bibfnamefont {V.~K.}\ \bibnamefont
  {Kalevich}}\ and\ \bibinfo {author} {\bibfnamefont {V.~G.}\ \bibnamefont
  {Fleisher}},\ }\bibfield  {title} {\bibinfo {title} {Optical detection of
  {NMR} with dynamic cooling of the nuclear spin system of a semiconductor by
  polarized light},\ }\href@noop {} {\bibfield  {journal} {\bibinfo  {journal}
  {Bull. Acad. Sci. USSR Phys. Ser.}\ }\textbf {\bibinfo {volume} {47}},\
  \bibinfo {pages} {5} (\bibinfo {year} {1983})}\BibitemShut {NoStop}%
\bibitem [{\citenamefont {Griffiths}\ \emph {et~al.}(2019)\citenamefont
  {Griffiths}, \citenamefont {Huang}, \citenamefont {Rastelli}, \citenamefont
  {Skolnick},\ and\ \citenamefont {Chekhovich}}]{Griffiths2019}%
  \BibitemOpen
  \bibfield  {author} {\bibinfo {author} {\bibfnamefont {I.~M.}\ \bibnamefont
  {Griffiths}}, \bibinfo {author} {\bibfnamefont {H.}~\bibnamefont {Huang}},
  \bibinfo {author} {\bibfnamefont {A.}~\bibnamefont {Rastelli}}, \bibinfo
  {author} {\bibfnamefont {M.~S.}\ \bibnamefont {Skolnick}},\ and\ \bibinfo
  {author} {\bibfnamefont {E.~A.}\ \bibnamefont {Chekhovich}},\ }\bibfield
  {title} {{\selectlanguage {English}\bibinfo {title} {{Complete
  characterization of GaAs gradient-elastic tensors and reconstruction of
  internal strain in GaAs/AlGaAs quantum dots using nuclear magnetic
  resonance}}},\ }\href {https://doi.org/10.1103/PhysRevB.99.125304} {\bibfield
   {journal} {\bibinfo  {journal} {Phys. Rev. B}\ }\textbf {\bibinfo {volume}
  {99}},\ \bibinfo {pages} {125304} (\bibinfo {year} {2019})}\BibitemShut
  {NoStop}%
\bibitem [{\citenamefont {Sobolev}\ \emph {et~al.}(1989)\citenamefont
  {Sobolev}, \citenamefont {Brunkov}, \citenamefont {Konnikov}, \citenamefont
  {Stepanova}, \citenamefont {Nikitin}, \citenamefont {Ulin}, \citenamefont
  {Dolbaya}, \citenamefont {Kamushadze},\ and\ \citenamefont
  {Ma\u{i}suradze}}]{Sem89}%
  \BibitemOpen
  \bibfield  {author} {\bibinfo {author} {\bibfnamefont {M.~M.}\ \bibnamefont
  {Sobolev}}, \bibinfo {author} {\bibfnamefont {P.~N.}\ \bibnamefont
  {Brunkov}}, \bibinfo {author} {\bibfnamefont {S.~G.}\ \bibnamefont
  {Konnikov}}, \bibinfo {author} {\bibfnamefont {M.~N.}\ \bibnamefont
  {Stepanova}}, \bibinfo {author} {\bibfnamefont {V.~G.}\ \bibnamefont
  {Nikitin}}, \bibinfo {author} {\bibfnamefont {V.~P.}\ \bibnamefont {Ulin}},
  \bibinfo {author} {\bibfnamefont {A.~S.}\ \bibnamefont {Dolbaya}}, \bibinfo
  {author} {\bibfnamefont {T.~D.}\ \bibnamefont {Kamushadze}},\ and\ \bibinfo
  {author} {\bibfnamefont {R.~M.}\ \bibnamefont {Ma\u{i}suradze}},\ }\bibfield
  {title} {\bibinfo {title} {Mechanism of compensation of multilayed structures
  made of undoped {GaAs} grown from molten solution in {Ga}},\ }\href@noop {}
  {\bibfield  {journal} {\bibinfo  {journal} {Sov. Phys. Semicond.}\ }\textbf
  {\bibinfo {volume} {23}},\ \bibinfo {pages} {660} (\bibinfo {year}
  {1989})}\BibitemShut {NoStop}%
\bibitem [{\citenamefont {Dzhioev}\ \emph {et~al.}(2002)\citenamefont
  {Dzhioev}, \citenamefont {Kavokin}, \citenamefont {Korenev}, \citenamefont
  {Lazarev}, \citenamefont {Meltser}, \citenamefont {Stepanova}, \citenamefont
  {Zakharchenya}, \citenamefont {Gammon},\ and\ \citenamefont
  {Katzer}}]{PhysRevB.66.245204}%
  \BibitemOpen
  \bibfield  {author} {\bibinfo {author} {\bibfnamefont {R.~I.}\ \bibnamefont
  {Dzhioev}}, \bibinfo {author} {\bibfnamefont {K.~V.}\ \bibnamefont
  {Kavokin}}, \bibinfo {author} {\bibfnamefont {V.~L.}\ \bibnamefont
  {Korenev}}, \bibinfo {author} {\bibfnamefont {M.~V.}\ \bibnamefont
  {Lazarev}}, \bibinfo {author} {\bibfnamefont {B.~Y.}\ \bibnamefont
  {Meltser}}, \bibinfo {author} {\bibfnamefont {M.~N.}\ \bibnamefont
  {Stepanova}}, \bibinfo {author} {\bibfnamefont {B.~P.}\ \bibnamefont
  {Zakharchenya}}, \bibinfo {author} {\bibfnamefont {D.}~\bibnamefont
  {Gammon}},\ and\ \bibinfo {author} {\bibfnamefont {D.~S.}\ \bibnamefont
  {Katzer}},\ }\bibfield  {title} {\bibinfo {title} {Low-temperature spin
  relaxation in $n$-type {GaAs}},\ }\href
  {https://doi.org/10.1103/PhysRevB.66.245204} {\bibfield  {journal} {\bibinfo
  {journal} {Phys. Rev. B}\ }\textbf {\bibinfo {volume} {66}},\ \bibinfo
  {pages} {245204} (\bibinfo {year} {2002})}\BibitemShut {NoStop}%
\bibitem [{\citenamefont {Paget}(1982)}]{PhysRevB.25.4444}%
  \BibitemOpen
  \bibfield  {author} {\bibinfo {author} {\bibfnamefont {D.}~\bibnamefont
  {Paget}},\ }\bibfield  {title} {\bibinfo {title} {Optical detection of {NMR}
  in high-purity {GaAs}: Direct study of the relaxation of nuclei close to
  shallow donors},\ }\href {https://doi.org/10.1103/PhysRevB.25.4444}
  {\bibfield  {journal} {\bibinfo  {journal} {Phys. Rev. B}\ }\textbf {\bibinfo
  {volume} {25}},\ \bibinfo {pages} {4444} (\bibinfo {year}
  {1982})}\BibitemShut {NoStop}%
\bibitem [{Note1()}]{Note1}%
  \BibitemOpen
  \bibinfo {note} {Assuming $D=10^{13}$~cm$^2$/s \cite {PhysRevB.25.4444},
  pumping time $t_p=60$~s and $l_d=2\protect \sqrt {D t_p}$}\BibitemShut
  {NoStop}%
\bibitem [{\citenamefont {Abragam}(1961)}]{Abragam}%
  \BibitemOpen
  \bibfield  {author} {\bibinfo {author} {\bibfnamefont {A.}~\bibnamefont
  {Abragam}},\ }\href@noop {} {\emph {\bibinfo {title} {The principles of
  nuclear magnetism}}}\ (\bibinfo  {publisher} {Clarendon Press},\ \bibinfo
  {year} {1961})\BibitemShut {NoStop}%
\bibitem [{Note2()}]{Note2}%
  \BibitemOpen
  \bibinfo {note} {In bulk GaAs crystals, the Overhauser field under optical
  pumping can reach several Tesla. We deliberately use weak measurement fields
  and moderate nuclear spin temperatures to match the Overhauser field at the
  measurement stage with the width of the electron Hanle effect in our
  sample}\BibitemShut {NoStop}%
\bibitem [{Note3()}]{Note3}%
  \BibitemOpen
  \bibinfo {note} {The intensity of the incident field is proportional to
  $B_1^2$}\BibitemShut {NoStop}%
\bibitem [{Note4()}]{Note4}%
  \BibitemOpen
  \bibinfo {note} {The consistent theory of the Hanle effect in presence of
  dynamic polarization of nuclear spins (including that in the Knight field of
  photoexcited electrons) is given in \cite {OO1}. It says that the
  magnetization of cooled nuclei in the Knight field does not affect the Hanle
  curve, since the resulted Overhauser field is parallel to the mean electron
  spin. As a consequence, the only contribution of the Knight field comes from
  additional cooling of the NSS, proportional to ${S_0}{S_z}$, where $S_0$ is
  the initial mean spin of photoexcited electrons at birth, and $S_z$ is their
  current spin projection on the excitation axis. Our experiments were designed
  so that the variation of the nuclear spin temperature due to cooling in the
  Knight field was much smaller than that due to preliminary optical cooling.
  It is only important at the very end of the spin-lattice relaxation of the
  NSS, as seen from Fig.~\ref {fig:timing}~(b). At this stage, the relative
  variation of $S_z$ (proportional to the PL polarization $\rho \left (t\right
  )$) is small, which justifies using a linearized rate equation of dynamic
  polarization, leading to the single-exponential time dependence of the
  Overhauser field.}\BibitemShut {Stop}%
\bibitem [{\citenamefont {Yamada}(1985)}]{Yamada1998l}%
  \BibitemOpen
  \bibfield  {author} {\bibinfo {author} {\bibfnamefont {M.}~\bibnamefont
  {Yamada}},\ }\bibfield  {title} {\bibinfo {title} {{Quantitative photoelastic
  measurement of residual strains in undoped semi-insulating gallium
  arsenide}},\ }\href@noop {} {\bibfield  {journal} {\bibinfo  {journal}
  {Applied Physics Letters}\ }\textbf {\bibinfo {volume} {47}},\ \bibinfo
  {pages} {365} (\bibinfo {year} {1985})}\BibitemShut {NoStop}%
\bibitem [{\citenamefont {Fitzsimmons}\ \emph {et~al.}(2007)\citenamefont
  {Fitzsimmons}, \citenamefont {Kirby}, \citenamefont {Hengartner},
  \citenamefont {Trouw}, \citenamefont {Erickson}, \citenamefont {Flexner},
  \citenamefont {Kondo}, \citenamefont {Adelmann}, \citenamefont
  {Palmstr\o{}m}, \citenamefont {Crowell}, \citenamefont {Chen}, \citenamefont
  {Gentile}, \citenamefont {Borchers}, \citenamefont {Majkrzak},\ and\
  \citenamefont {Pynn}}]{PhysRevB.76.245301}%
  \BibitemOpen
  \bibfield  {author} {\bibinfo {author} {\bibfnamefont {M.~R.}\ \bibnamefont
  {Fitzsimmons}}, \bibinfo {author} {\bibfnamefont {B.~J.}\ \bibnamefont
  {Kirby}}, \bibinfo {author} {\bibfnamefont {N.~W.}\ \bibnamefont
  {Hengartner}}, \bibinfo {author} {\bibfnamefont {F.}~\bibnamefont {Trouw}},
  \bibinfo {author} {\bibfnamefont {M.~J.}\ \bibnamefont {Erickson}}, \bibinfo
  {author} {\bibfnamefont {S.~D.}\ \bibnamefont {Flexner}}, \bibinfo {author}
  {\bibfnamefont {T.}~\bibnamefont {Kondo}}, \bibinfo {author} {\bibfnamefont
  {C.}~\bibnamefont {Adelmann}}, \bibinfo {author} {\bibfnamefont {C.~J.}\
  \bibnamefont {Palmstr\o{}m}}, \bibinfo {author} {\bibfnamefont {P.~A.}\
  \bibnamefont {Crowell}}, \bibinfo {author} {\bibfnamefont {W.~C.}\
  \bibnamefont {Chen}}, \bibinfo {author} {\bibfnamefont {T.~R.}\ \bibnamefont
  {Gentile}}, \bibinfo {author} {\bibfnamefont {J.~A.}\ \bibnamefont
  {Borchers}}, \bibinfo {author} {\bibfnamefont {C.~F.}\ \bibnamefont
  {Majkrzak}},\ and\ \bibinfo {author} {\bibfnamefont {R.}~\bibnamefont
  {Pynn}},\ }\bibfield  {title} {\bibinfo {title} {Suppression of nuclear
  polarization near the surface of optically pumped {GaAs}},\ }\href
  {https://doi.org/10.1103/PhysRevB.76.245301} {\bibfield  {journal} {\bibinfo
  {journal} {Phys. Rev. B}\ }\textbf {\bibinfo {volume} {76}},\ \bibinfo
  {pages} {245301} (\bibinfo {year} {2007})}\BibitemShut {NoStop}%
\bibitem [{\citenamefont {King}\ \emph {et~al.}(2012)\citenamefont {King},
  \citenamefont {Li}, \citenamefont {Meriles},\ and\ \citenamefont
  {Reimer}}]{King2012}%
  \BibitemOpen
  \bibfield  {author} {\bibinfo {author} {\bibfnamefont {J.~P.}\ \bibnamefont
  {King}}, \bibinfo {author} {\bibfnamefont {Y.}~\bibnamefont {Li}}, \bibinfo
  {author} {\bibfnamefont {C.~A.}\ \bibnamefont {Meriles}},\ and\ \bibinfo
  {author} {\bibfnamefont {J.~A.}\ \bibnamefont {Reimer}},\ }\bibfield  {title}
  {\bibinfo {title} {Optically rewritable patterns of nuclear magnetization in
  gallium arsenide},\ }\href {https://doi.org/10.1038/ncomms1918} {\bibfield
  {journal} {\bibinfo  {journal} {Nat. Comm.}\ }\textbf {\bibinfo {volume}
  {3}},\ \bibinfo {pages} {918} (\bibinfo {year} {2012})}\BibitemShut {NoStop}%
\bibitem [{\citenamefont {Dumas}\ \emph {et~al.}(1979)\citenamefont {Dumas},
  \citenamefont {Soest}, \citenamefont {Sher},\ and\ \citenamefont
  {Swiggard}}]{PhysRevB.20.4406}%
  \BibitemOpen
  \bibfield  {author} {\bibinfo {author} {\bibfnamefont {K.~A.}\ \bibnamefont
  {Dumas}}, \bibinfo {author} {\bibfnamefont {J.~F.}\ \bibnamefont {Soest}},
  \bibinfo {author} {\bibfnamefont {A.}~\bibnamefont {Sher}},\ and\ \bibinfo
  {author} {\bibfnamefont {E.~M.}\ \bibnamefont {Swiggard}},\ }\bibfield
  {title} {\bibinfo {title} {Electrically induced shifts of the {GaAs} nuclear
  spin levels},\ }\href {https://doi.org/10.1103/PhysRevB.20.4406} {\bibfield
  {journal} {\bibinfo  {journal} {Phys. Rev. B}\ }\textbf {\bibinfo {volume}
  {20}},\ \bibinfo {pages} {4406} (\bibinfo {year} {1979})}\BibitemShut
  {NoStop}%
\bibitem [{\citenamefont {Chekhovich}\ \emph {et~al.}(2018)\citenamefont
  {Chekhovich}, \citenamefont {Griffiths}, \citenamefont {Skolnick},
  \citenamefont {Huang}, \citenamefont {da~Silva}, \citenamefont {Yuan},\ and\
  \citenamefont {Rastelli}}]{PhysRevB.97.235311}%
  \BibitemOpen
  \bibfield  {author} {\bibinfo {author} {\bibfnamefont {E.~A.}\ \bibnamefont
  {Chekhovich}}, \bibinfo {author} {\bibfnamefont {I.~M.}\ \bibnamefont
  {Griffiths}}, \bibinfo {author} {\bibfnamefont {M.~S.}\ \bibnamefont
  {Skolnick}}, \bibinfo {author} {\bibfnamefont {H.}~\bibnamefont {Huang}},
  \bibinfo {author} {\bibfnamefont {S.~F.~C.}\ \bibnamefont {da~Silva}},
  \bibinfo {author} {\bibfnamefont {X.}~\bibnamefont {Yuan}},\ and\ \bibinfo
  {author} {\bibfnamefont {A.}~\bibnamefont {Rastelli}},\ }\bibfield  {title}
  {\bibinfo {title} {Cross calibration of deformation potentials and
  gradient-elastic tensors of {GaAs} using photoluminescence and nuclear
  magnetic resonance spectroscopy in {GaAs/AlGaAs} quantum dot structures},\
  }\href {https://doi.org/10.1103/PhysRevB.97.235311} {\bibfield  {journal}
  {\bibinfo  {journal} {Phys. Rev. B}\ }\textbf {\bibinfo {volume} {97}},\
  \bibinfo {pages} {235311} (\bibinfo {year} {2018})}\BibitemShut {NoStop}%
\bibitem [{\citenamefont {Burenkov}\ \emph {et~al.}(1973)\citenamefont
  {Burenkov}, \citenamefont {Burdukov}, \citenamefont {Davidov},\ and\
  \citenamefont {Nikanorov}}]{SS73}%
  \BibitemOpen
  \bibfield  {author} {\bibinfo {author} {\bibfnamefont {Y.~A.}\ \bibnamefont
  {Burenkov}}, \bibinfo {author} {\bibfnamefont {Y.~M.}\ \bibnamefont
  {Burdukov}}, \bibinfo {author} {\bibfnamefont {S.~Y.}\ \bibnamefont
  {Davidov}},\ and\ \bibinfo {author} {\bibfnamefont {S.~P.}\ \bibnamefont
  {Nikanorov}},\ }\href@noop {} {\bibfield  {journal} {\bibinfo  {journal}
  {Sov. Phys. Solid State}\ }\textbf {\bibinfo {volume} {15}},\ \bibinfo
  {pages} {1175} (\bibinfo {year} {1973})}\BibitemShut {NoStop}%
\bibitem [{\citenamefont {Brun}\ \emph {et~al.}(1963)\citenamefont {Brun},
  \citenamefont {Mahler}, \citenamefont {Mahon},\ and\ \citenamefont
  {Pierce}}]{PhysRev.129.1965}%
  \BibitemOpen
  \bibfield  {author} {\bibinfo {author} {\bibfnamefont {E.}~\bibnamefont
  {Brun}}, \bibinfo {author} {\bibfnamefont {R.~J.}\ \bibnamefont {Mahler}},
  \bibinfo {author} {\bibfnamefont {H.}~\bibnamefont {Mahon}},\ and\ \bibinfo
  {author} {\bibfnamefont {W.~L.}\ \bibnamefont {Pierce}},\ }\bibfield  {title}
  {\bibinfo {title} {Electrically induced nuclear quadrupole spin transitions
  in a {GaAs} single crystal},\ }\href
  {https://doi.org/10.1103/PhysRev.129.1965} {\bibfield  {journal} {\bibinfo
  {journal} {Phys. Rev.}\ }\textbf {\bibinfo {volume} {129}},\ \bibinfo {pages}
  {1965} (\bibinfo {year} {1963})}\BibitemShut {NoStop}%
\bibitem [{\citenamefont {Pyykk\"{o}}(2008)}]{doi:10.1080/00268970802018367}%
  \BibitemOpen
  \bibfield  {author} {\bibinfo {author} {\bibfnamefont {P.}~\bibnamefont
  {Pyykk\"{o}}},\ }\bibfield  {title} {\bibinfo {title} {Year-2008 nuclear
  quadrupole moments},\ }\href@noop {} {\bibfield  {journal} {\bibinfo
  {journal} {Molecular Physics}\ }\textbf {\bibinfo {volume} {106}},\ \bibinfo
  {pages} {1965} (\bibinfo {year} {2008})}\BibitemShut {NoStop}%
\bibitem [{\citenamefont {Yin}\ \emph {et~al.}(1991)\citenamefont {Yin},
  \citenamefont {Yan}, \citenamefont {Pollak}, \citenamefont {Pettit},\ and\
  \citenamefont {Woodall}}]{PhysRevB.43.12138}%
  \BibitemOpen
  \bibfield  {author} {\bibinfo {author} {\bibfnamefont {Y.}~\bibnamefont
  {Yin}}, \bibinfo {author} {\bibfnamefont {D.}~\bibnamefont {Yan}}, \bibinfo
  {author} {\bibfnamefont {F.~H.}\ \bibnamefont {Pollak}}, \bibinfo {author}
  {\bibfnamefont {G.~D.}\ \bibnamefont {Pettit}},\ and\ \bibinfo {author}
  {\bibfnamefont {J.~M.}\ \bibnamefont {Woodall}},\ }\bibfield  {title}
  {\bibinfo {title} {Observation of franz-keldysh oscillations in the
  stress-modulated spectra of (001) $n$-type {GaAs}},\ }\href
  {https://doi.org/10.1103/PhysRevB.43.12138} {\bibfield  {journal} {\bibinfo
  {journal} {Phys. Rev. B}\ }\textbf {\bibinfo {volume} {43}},\ \bibinfo
  {pages} {12138} (\bibinfo {year} {1991})}\BibitemShut {NoStop}%
\bibitem [{\citenamefont {Frigeri}(2001)}]{FRIGERI20012557}%
  \BibitemOpen
  \bibfield  {author} {\bibinfo {author} {\bibfnamefont {C.}~\bibnamefont
  {Frigeri}},\ }\bibfield  {title} {\bibinfo {title} {{Electron Beam-induced
  Current}},\ }in\ \href
  {https://doi.org/https://doi.org/10.1016/B0-08-043152-6/00463-0} {\emph
  {\bibinfo {booktitle} {Encyclopedia of Materials: Science and Technology}}},\
  \bibinfo {editor} {edited by\ \bibinfo {editor} {\bibfnamefont {K.~H.~J.}\
  \bibnamefont {Buschow}}, \bibinfo {editor} {\bibfnamefont {R.~W.}\
  \bibnamefont {Cahn}}, \bibinfo {editor} {\bibfnamefont {M.~C.}\ \bibnamefont
  {Flemings}}, \bibinfo {editor} {\bibfnamefont {B.}~\bibnamefont {Ilschner}},
  \bibinfo {editor} {\bibfnamefont {E.~J.}\ \bibnamefont {Kramer}}, \bibinfo
  {editor} {\bibfnamefont {S.}~\bibnamefont {Mahajan}},\ and\ \bibinfo {editor}
  {\bibfnamefont {P.}~\bibnamefont {Veyssi{\`e}re}}}\ (\bibinfo  {publisher}
  {Elsevier},\ \bibinfo {address} {Oxford},\ \bibinfo {year} {2001})\ pp.\
  \bibinfo {pages} {2557--2563}\BibitemShut {NoStop}%
\bibitem [{\citenamefont {Oelgart}\ \emph {et~al.}(1981)\citenamefont
  {Oelgart}, \citenamefont {Fiddicke},\ and\ \citenamefont {Reulke}}]{Oelgart}%
  \BibitemOpen
  \bibfield  {author} {\bibinfo {author} {\bibfnamefont {G.}~\bibnamefont
  {Oelgart}}, \bibinfo {author} {\bibfnamefont {J.}~\bibnamefont {Fiddicke}},\
  and\ \bibinfo {author} {\bibfnamefont {R.}~\bibnamefont {Reulke}},\
  }\bibfield  {title} {\bibinfo {title} {{Investigation of minority-carrier
  diffusion lengths by means of the scanning electron microprobe (SEM)}},\
  }\href {https://doi.org/10.1002/pssa.2210660135} {\bibfield  {journal}
  {\bibinfo  {journal} {Phys. Status Solidi (a)}\ }\textbf {\bibinfo {volume}
  {66}},\ \bibinfo {pages} {283} (\bibinfo {year} {1981})}\BibitemShut
  {NoStop}%
\bibitem [{\citenamefont {L{\"u}th}(2015)}]{Luth2015}%
  \BibitemOpen
  \bibfield  {author} {\bibinfo {author} {\bibfnamefont {H.}~\bibnamefont
  {L{\"u}th}},\ }\bibinfo {title} {Space-charge layers at semiconductor
  interfaces},\ in\ \href {https://doi.org/10.1007/978-3-319-10756-1_7} {\emph
  {\bibinfo {booktitle} {Solid Surfaces, Interfaces and Thin Films}}}\
  (\bibinfo  {publisher} {Springer International Publishing},\ \bibinfo
  {address} {Cham},\ \bibinfo {year} {2015})\ pp.\ \bibinfo {pages}
  {337--391}\BibitemShut {NoStop}%
\end{thebibliography}%

\end{document}